\documentclass[fleqn,usenatbib]{mnras}

\usepackage{newtxtext,newtxmath}

\usepackage[T1]{fontenc}
\usepackage{multirow} 
\usepackage{ragged2e}
\DeclareRobustCommand{\VAN}[3]{#2}
\let\VANthebibliography\thebibliography
\def\thebibliography{\DeclareRobustCommand{\VAN}[3]{##3}\VANthebibliography}

\usepackage{graphicx}	
\usepackage{amsmath}	
\graphicspath{{./}{plots/}}
\usepackage{tabularx}
\usepackage{caption}

\usepackage{float}
\usepackage{enumitem}
\usepackage{graphicx}	
\usepackage{amsmath}	

\usepackage{amssymb}	
\usepackage{natbib}
\usepackage{threeparttable}
\usepackage{color}

\usepackage{lineno}

\title{Testing Spin Prior Assumptions of Binary Black Hole Parameter Inference and Their Astrophysical Implications}
\author[Chen-He\, Wu et al.]{Chen-He\, Wu,$^{1}$
Wei-Hua\, Guo,$^{2}$\thanks{E-mail: whg038275@gmail.com}
Shu-Jin\, Hou,$^{2}$
Yuan-Zhu\, Wang,$^{3}$
Ying\, Qin,$^{4,5}$ \\
$^{1}$Lester B. Pearson United World College of the Pacific, Victoria, British Columbia, V9C 4H7, Canada\\
$^{2}$Department of Physics and Electronic Engineering, Nanyang Normal University, Nanyang, Henan, 473061, China\\
$^{3}$Institute for Theoretical Physics and Cosmology, Zhejiang University of Technology, Hangzhou,
310032, China\\
$^{4}$Department of Physics, Anhui Normal University, Wuhu, Anhui, 241002, China\\
$^{5}$Center for Astrophysics and Astronomical Technology, Anhui Normal University, Wuhu, Anhui, 241002, China
}

\begin{document}
\label{firstpage}
\pagerange{\pageref{firstpage}--\pageref{lastpage}}
\maketitle

\begin{abstract}
With the growing number of binary black hole (BBH) candidates in the Gravitational-Wave Transient Catalog 5.0 (GWTC-5.0), an increasing number of BBH systems with distinct properties have been reported. The individual spins are relatively less constrained in the parameter estimations for most events. While the LIGO–Virgo–KAGRA Collaboration (LVK) adopts an uninformative spin prior by default, some binary evolution scenarios have strong predictions on component spins. In this work, we perform Bayesian parameter inference for a subset of BBH events spanning a broad range of properties in the chirp mass, mass ratio, and effective inspiral spin ($\mathcal{M}_{\rm c}$-$q$-$\chi_{\rm eff}$) parameter space. We consider three different spin priors: the LVK spin prior and two astrophysically motivated priors, assuming a non-spinning primary BH ($\chi_1 = 0$) or a non-spinning secondary BH ($\chi_2 = 0$). We find that the non-spinning primary BH prior is disfavored by the data for several systems with high effective inspiral spins, including GW190517\_055101, GW190412, GW241113\_163507, and GW231028\_153006. We suggest that standard common-envelope evolution represents a promising formation channel that can be tested with gravitational-wave observations, and propose a novel method for estimating the merger rate of systems formed through alternative evolutionary channels. We further suggest that the binary evolution involving mass-ratio reversal may provide a possible formation pathway for these systems.
\end{abstract}




\section{Introduction}\label{sect1}
Recent gravitational-wave observations have substantially expanded the observed population of binary black hole (BBH) mergers, providing increasingly detailed insights into the masses and spins of BHs \citep{gwtc5_pop}. In particular, the Gravitational-Wave Transient Catalog 5.0 (GWTC-5.0) \citep{gwtc5_catalog} has revealed an increasing number of BBH systems with rapidly spinning BHs. The inferred properties of BBH systems, including their component masses and spin parameters, exhibit a wide diversity, yet their formation channels remain uncertain.

Although the individual BH spins can provide valuable clues to the formation and evolutionary history of BBH systems, they remain poorly constrained for many events, and their inferred posterior distributions are sensitive to the assumed priors. The LIGO–Virgo–KAGRA Collaboration (LVK) adopts a relatively uninformative spin prior, in which spin magnitudes are distributed uniformly and spin orientations are assumed to be isotropic. In isolated binary evolution, the primary BH (massive component) is expected to form first and possess a negligible spin under the commonly adopted assumption of efficient angular-momentum transport within its progenitor \citep{Qin2018,fuller2019}. Motivated by this astrophysical expectation, several studies have reanalyzed the properties of BBH systems, including GW190412 \citep{Mandel2020} and GW190814 \citep{GW190814}, as well as black hole–neutron star binaries \citep{Mandel2021,Xue2025}.

With the high effective spins inferred for several BBH mergers \citep{Abbott2021,Abbott2024}, \citet{Olejak2021} proposed an isolated-binary formation channel capable of producing rapidly spinning primary BHs. In this scenario, the initially less massive star accretes substantial mass from its companion during the first mass-transfer phase, eventually becoming the more massive component through mass-ratio reversal (MRR) and subsequently forming a second-born BH with a spin magnitude of 0.68. \citet{Broekgaarden2022} carried out detailed population-synthesis studies of BBHs formed through this channel and found the MRR scenario could be a common evolutionary outcome in BBH formation. More recently, this channel has been proposed as a possible formation pathway for GW241011\_233834 \citep{Hu2026}.

In this work, we investigate how astrophysically motivated spin priors affect the posterior distributions and Bayesian evidences for a subset of BBH events spanning a broad range of properties in the chirp mass, mass ratio, and effective inspiral spin ($\mathcal{M}_{\rm c}$-$q$-$\chi_{\rm eff}$) parameter space. We consider three spin-prior configurations: the default LVK uniform spin prior, a non-spinning primary BH ($\chi_1 = 0$) prior, and a non-spinning secondary BH ($\chi_2 = 0$) prior. We describe the event selection and adopted spin priors in Section~\ref{sect2}, present the parameter-estimation analysis and results in Section~\ref{sect3}, and summarize our main findings and discuss their potential implications in Section~\ref{sect4}.

\section{Sample Selection and Spin Prior Assumptions}\label{sect2}
\subsection{Sample Selection}
\begin{figure*}
    \centering
\includegraphics[width=0.95\textwidth]{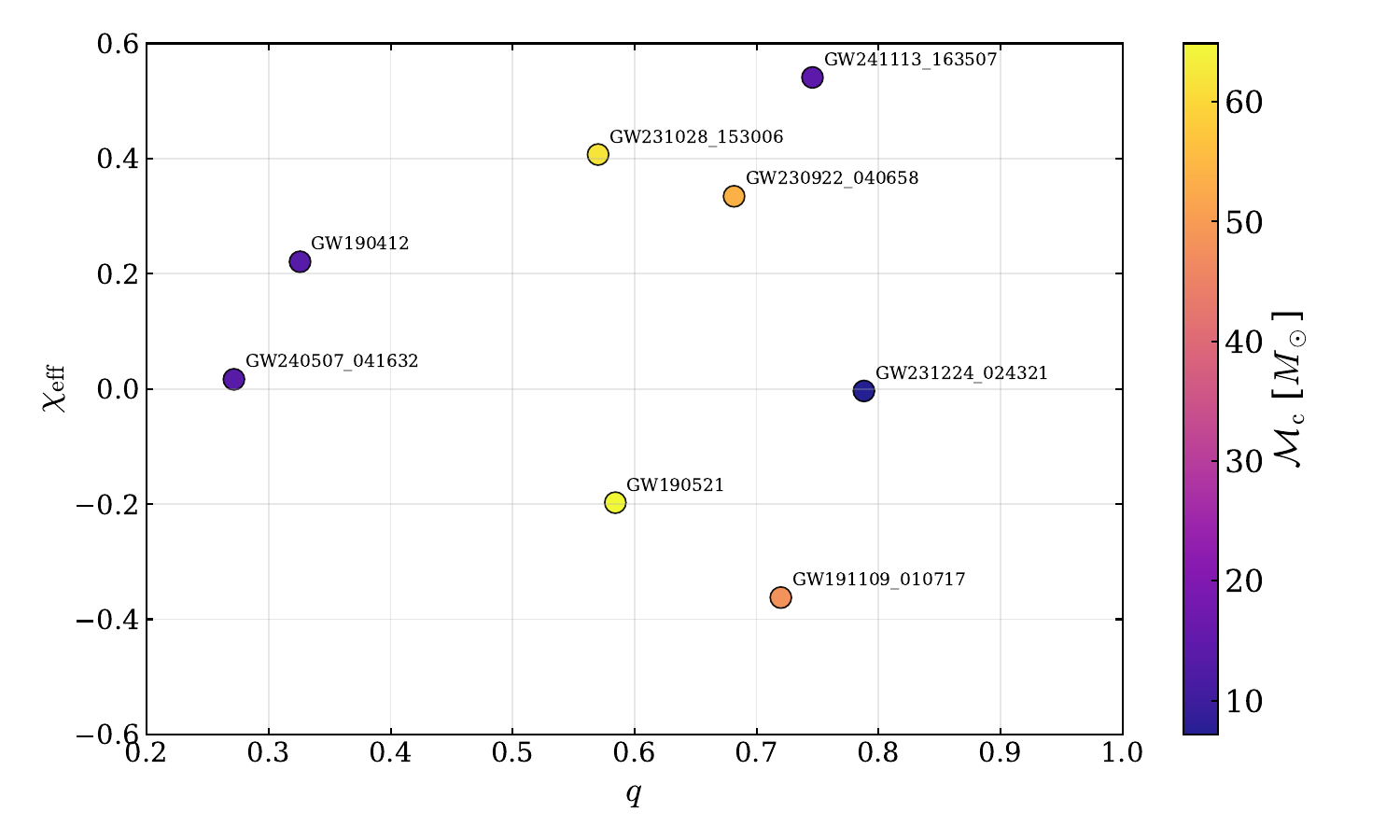}
    \caption{Selected events and their inferred posterior medians in the $\chi_{\rm eff}$-$q$ plane. The color of each event represents its medians of chirp mass, $\mathcal{M}_{\rm c}$.} 
    \label{fig:samples}
\end{figure*}

We select a subset of eight events that span a wide range of properties in the parameter space of $\mathcal{M}_{\rm c}$-$q$-$\chi_{\rm eff}$. We exclude GW241011\_233834, GW241110\_124123 and GW231123\_135430 because these events have been suggested to be more likely to originate from dynamical formation channels \citep{GW231123,GW241011}. For each selected event, we adopt the median values of $\mathcal{M}_{\rm c}$, $q$, and $\chi_{\rm eff}$ from their respective posterior distributions as point estimates. Each parameter is then normalized to the range of $[0,1]$ using its minimum and maximum values across the selected events, denoted by $\tilde{\mathcal{M}_{\rm c}}$, $\tilde{q}$, and $\tilde{\chi_{\rm eff}}$, respectively. We then define a ``length" parameter,  $\mathbf{L}$, to quantify how extreme each event is within the normalized $\tilde{\mathcal{M}_{\rm c}}$- $\tilde{q}$-$\tilde{\chi_{\rm eff}}$ parameter space: 

\begin{equation}\label{eq:length}
    \mathbf{L}=d_1\cdot\tilde{\mathcal{M}_{\rm c}} + d_2\cdot\tilde{q}+d_3\cdot\tilde{\chi_{\rm eff}},
\end{equation}
where the vector $(d_1,d_2,d_3) = (\pm1,\pm1,\pm1)$ specifies whether the high or low value of each dimension is desired. For example, to identify a source with large $\mathcal{M}_{\rm c}$, small $q$ and small $\chi_{\rm eff}$, we set $(d_1,d_2,d_3) = (+1,-1,-1)$ and select the event with the largest $\mathbf{L}$. Applying these criteria, we identify eight events with the largest values of $\mathbf{L}$ according to E.q.(\ref{eq:length}), as shown in Figure~\ref{fig:samples}.

\subsection{Spin Prior Assumptions}

In order to investigate the dependence of the inferred parameters on the adopted spin prior, we perform Bayesian inference under three different spin-prior configurations, as described below.

\subsubsection{LVK prior}
The dimensionless spin magnitudes of both BHs are assumed to be uniformly distributed, while the spin tilt angles are isotropically distributed. Specifically, we adopt a uniform prior on the spin magnitudes, $\chi_i \sim \mathcal{U}(0,1)$, and an isotropic distribution for the spin orientations, $\cos\theta_i \sim \mathcal{U}(-1,1)$, for both components. This configuration corresponds to the default spin prior commonly adopted in LVK parameter estimation analyses \citep{Abbott2020}.

\subsubsection{$\chi_1=0$ prior}
We assume zero spin magnitude for the primary BH, as expected in the standard common-envelope channel \citep[e.g.,][]{Qin2018,fuller2019}. The spin magnitude of the secondary BH is allowed to span the full physically permitted range, from zero to the maximal value, reflecting the competing effects of tidal spin-up and stellar-wind mass loss \citep{Qin2018,Bavera2020}. We further assume that the spin orientation of the secondary BH is isotropic, with $\cos\theta_2 \sim \mathcal{U}(-1,1)$.

\subsubsection{$\chi_2=0$ prior}
 In addition to the LVK prior and the $\chi_1=0$ prior, we also adopt an alternative spin prior, namely $\chi_2=0$ prior. In this scenario, the initially more massive star evolves faster and initiates mass transfer onto its companion through the first Roche-lobe overflow. It subsequently loses most of its hydrogen-rich envelope and becomes a helium star, provided that its mass is sufficiently high, and eventually forms the less massive, first-born BH (the secondary BH). The spin of this BH is expected to be negligible because its progenitor loses substantial angular momentum as a result of strong core–envelope coupling after leaving the main sequence \citep{Qin2018,fuller2019}. In contrast, the initially less massive star accretes sufficient mass during the first mass-transfer episode to reverse the mass ratio, and subsequently evolves into the more massive, second-born BH (i.e., the primary BH) following the common-envelope phase. Its natal spin can be relatively high because its progenitor may undergo efficient tidal spin-up during the late stages of binary evolution. This evolutionary pathway therefore provides an alternative formation scenario for GW241011\_233834 \citep{Hu2026}. We also assume an isotropic spin orientation for the primary BH, such that $\cos\theta_2 \sim \mathcal{U}(-1,1)$. 

\section{Parameter estimation and Results}\label{sect3}
\subsection{Parameter estimation}

We perform parameter estimation and spin-prior comparisons within the Bayesian inference framework. For each event, the publicly available strain data sampled at 4096 Hz are adopted. In our analysis, we employ the \texttt{IMRPhenomXPHM} waveform model \citep{Colleoni2025}, a frequency-domain phenomenological model describing the inspiral, merger, and ringdown phases of binary black hole coalescence. It incorporates spin-precession effects through the twisting-up approach and includes higher-order modes beyond the dominant quadrupole contribution. Owing to its balance between computational efficiency and modeling accuracy, \texttt{IMRPhenomXPHM} is widely used in computationally intensive gravitational-wave data analysis applications.

The posterior distributions of the binary parameters are obtained from the observed strain data, the adopted waveform model, the noise power spectral density (PSD), and the prior distributions. The Bayesian evidence, $\mathcal{Z}$, for a given model $\mathcal{M}$ is defined as 
\begin{equation}
    \mathcal{Z}\equiv p(d|\mathcal{M})=\int d\vartheta\mathcal{L}(d|\vartheta, \mathcal{M})\pi(\vartheta|\mathcal{M}),
\end{equation}
where $\vartheta$ denotes the set of parameters (e.g., spins, masses), $\mathcal{L}$ is the likelihood of the observed data given these parameters, and $\pi$ is the prior probability density assigned to the parameters in the model. 

Following the same approach \citep{Zevin2020,Qin2022}, we derive the Bayes factor, $\mathcal{B}$, between two models by comparing their Bayesian evidences \citep{Thrane2019,Zevin2020}. In our case, the two models differ only in their adopted spin priors, while the waveform model is kept identical. The Bayes factor between the LVK prior and the $\chi_i=0$ prior is therefore given by
\begin{equation}
    \mathcal{B}_{\mathrm{LVK}/\chi_i=0}=\frac{\mathcal{Z}_{\rm LVK}}{\mathcal{Z}_{\chi_i=0}}.
\end{equation}
The Bayes factor quantifies the relative statistical support for the two prior configurations provided by the data, while accounting for both the agreement between the model and the data and the prior volume. Thus, for each event, a larger Bayes factor indicates stronger relative support for the LVK prior. For convenience, we report the natural logarithmic Bayes factor, $\rm ln\mathcal{B}$,

\begin{equation}
    \ln\mathcal{B}_{\mathrm{LVK}/\chi_i=0}=
    \mathrm{ln}\mathcal{Z}_{\rm LVK}-\mathrm{ln}\mathcal{Z}_{\chi_i=0}.
\end{equation}
A positive value indicates stronger support for the LVK priors, whereas a negative value favors the $\chi_i=0$ model. 

We use the publicly released gravitational-wave strain data from the Gravitational Wave Open Science Center. Parameter estimation is performed using the Python package \textit{bilby} and the nested sampler \textit{nessai}. The event-specific analysis settings, including duration, sampling frequency, frequency bounds, and PSD setup are adopted from the corresponding publicly released parameter-estimation data files.

\begin{table*}
\centering
\caption{Posterior parameter estimates and Bayes factors for the BBH events under the LVK, $\chi_1=0$, and $\chi_2=0$ priors, where $q=m_2/m_1$. Values are medians with 90\% credible intervals.}
\label{table1}
\scriptsize
\setlength{\tabcolsep}{2.5pt}
\renewcommand{\arraystretch}{2.0}

\resizebox{\textwidth}{!}{%
\begin{tabular}{llcccccccccc}
\hline
Name & Prior & $m_1\,[M_\odot]$ & $m_2\,[M_\odot]$ & $q$ &
$\chi_1$ & $\chi_2$ & $\cos\theta_1$ & $\cos\theta_2$ &
$\chi_{\mathrm{eff}}$ &
$\ln\mathcal{B}_{\mathrm{LVK}/\chi_1=0}$ &
$\ln\mathcal{B}_{\mathrm{LVK}/\chi_2=0}$ \\
\hline

\multicolumn{12}{c}{\textbf{Group I}} \\
\hline

  &$\chi_1=0$ &
$29.23_{-4.70}^{+4.97}$ &
$25.93_{-5.59}^{+4.78}$ &
$0.91_{-0.22}^{+0.08}$ &
$0$ &
$0.92_{-0.20}^{+0.06}$ &
$1$ &
$0.89_{-0.26}^{+0.10}$ &
$0.37_{-0.14}^{+0.08}$ &
$6.04$ &
/ \\

GW190517\_055101 & $\chi_2=0$ &
$39.92_{-7.74}^{+9.72}$ &
$21.90_{-4.71}^{+5.10}$ &
$0.55_{-0.18}^{+0.20}$ &
$0.93_{-0.14}^{+0.06}$ &
$0$ &
$0.79_{-0.30}^{+0.18}$ &
$1$ &
$0.45_{-0.15}^{+0.11}$ &
/ &
$0.67$ \\

& LVK &
$36.38_{-7.37}^{+8.83}$ &
$23.06_{-5.14}^{+5.73}$ &
$0.64_{-0.21}^{+0.26}$ &
$0.91_{-0.21}^{+0.08}$ &
$0.62_{-0.51}^{+0.33}$ &
$0.81_{-0.27}^{+0.16}$ &
$0.31_{-0.93}^{+0.58}$ &
$0.50_{-0.18}^{+0.17}$ &
/ &
/ \\

\hline

& $\chi_1=0$ &
$24.99_{-3.00}^{+1.90}$ &
$9.94_{-0.73}^{+1.29}$ &
$0.40_{-0.05}^{+0.11}$ &
$0$ &
$0.85_{-0.38}^{+0.13}$ &
$1$ &
$0.78_{-0.44}^{+0.20}$ &
$0.18_{-0.11}^{+0.06}$ &
$4.81$ &
/ \\

GW190412 & $\chi_2=0$ &
$30.14_{-3.90}^{+4.53}$ &
$8.39_{-0.96}^{+1.09}$ &
$0.28_{-0.06}^{+0.08}$ &
$0.38_{-0.11}^{+0.12}$ &
$0$ &
$0.84_{-0.21}^{+0.13}$ &
$1$ &
$0.24_{-0.08}^{+0.08}$ &
/ &
$0.92$ \\

& LVK &
$30.15_{-3.58}^{+4.53}$ &
$8.40_{-0.92}^{+1.02}$ &
$0.28_{-0.06}^{+0.07}$ &
$0.36_{-0.13}^{+0.12}$ &
$0.48_{-0.42}^{+0.44}$ &
$0.81_{-0.28}^{+0.14}$ &
$0.28_{-0.97}^{+0.62}$ &
$0.25_{-0.09}^{+0.09}$ &
/ &
/ \\

\hline

& $\chi_1=0$ &
$16.91_{-1.53}^{+1.62}$ &
$15.77_{-1.65}^{+1.68}$ &
$0.95_{-0.12}^{+0.05}$ &
$0$ &
$0.94_{-0.12}^{+0.04}$ &
$1$ &
$0.95_{-0.13}^{+0.04}$ &
$0.43_{-0.08}^{+0.04}$ &
$3.30$ &
/ \\

GW241113\_163507 & $\chi_2=0$ &
$19.55_{-2.83}^{+4.48}$ &
$14.11_{-2.85}^{+2.35}$ &
$0.73_{-0.24}^{+0.21}$ &
$0.89_{-0.16}^{+0.09}$ &
$0$ &
$0.93_{-0.16}^{+0.06}$ &
$1$ &
$0.47_{-0.08}^{+0.07}$ &
/ &
$-0.08$ \\

& LVK &
$18.74_{-2.45}^{+4.16}$ &
$14.26_{-2.81}^{+2.32}$ &
$0.77_{-0.25}^{+0.19}$ &
$0.80_{-0.29}^{+0.16}$ &
$0.50_{-0.41}^{+0.40}$ &
$0.82_{-0.33}^{+0.15}$ &
$0.62_{-0.91}^{+0.33}$ &
$0.48_{-0.10}^{+0.09}$ &
/ &
/ \\

\hline

& $\chi_1=0$ &
$77.42_{-6.76}^{+10.25}$ &
$70.07_{-9.06}^{+9.04}$ &
$0.92_{-0.16}^{+0.08}$ &
$0$ &
$0.93_{-0.15}^{+0.06}$ &
$1$ &
$0.93_{-0.18}^{+0.06}$ &
$0.40_{-0.10}^{+0.06}$ &
$2.80$ &
/ \\

GW231028\_153006 & $\chi_2=0$ &
$89.20_{-13.15}^{+15.89}$ &
$60.74_{-12.67}^{+11.42}$ &
$0.68_{-0.21}^{+0.23}$ &
$0.90_{-0.18}^{+0.08}$ &
$0$ &
$0.86_{-0.25}^{+0.13}$ &
$1$ &
$0.44_{-0.11}^{+0.08}$ &
/ &
$0.37$ \\

& LVK &
$84.91_{-10.21}^{+12.37}$ &
$64.66_{-10.73}^{+10.33}$ &
$0.77_{-0.19}^{+0.18}$ &
$0.83_{-0.37}^{+0.14}$ &
$0.54_{-0.44}^{+0.37}$ &
$0.85_{-0.32}^{+0.13}$ &
$0.51_{-0.81}^{+0.41}$ &
$0.49_{-0.13}^{+0.11}$ &
/ &
/ \\

\hline

\multicolumn{12}{c}{\textbf{Group II}} \\
\hline

& $\chi_1=0$ &
$77.10_{-19.56}^{+34.50}$ &
$50.55_{-37.73}^{+22.46}$ &
$0.73_{-0.61}^{+0.24}$ &
$0$ &
$0.80_{-0.51}^{+0.17}$ &
$1$ &
$0.78_{-0.96}^{+0.20}$ &
$0.22_{-0.23}^{+0.19}$ &
$2.28$ &
/ \\

GW230922\_040658 & $\chi_2=0$ &
$84.48_{-21.76}^{+22.81}$ &
$37.06_{-16.40}^{+25.16}$ &
$0.45_{-0.24}^{+0.42}$ &
$0.70_{-0.32}^{+0.25}$ &
$0$ &
$0.82_{-0.44}^{+0.16}$ &
$1$ &
$0.37_{-0.24}^{+0.16}$ &
/ &
$0.63$ \\

& LVK &
$82.74_{-21.10}^{+23.16}$ &
$40.32_{-17.24}^{+24.94}$ &
$0.50_{-0.26}^{+0.42}$ &
$0.64_{-0.32}^{+0.29}$ &
$0.62_{-0.48}^{+0.33}$ &
$0.77_{-0.48}^{+0.20}$ &
$0.57_{-0.95}^{+0.38}$ &
$0.42_{-0.22}^{+0.19}$ &
/ &
/ \\

\hline

& $\chi_1=0$ &
$85.29_{-13.30}^{+14.91}$ &
$59.77_{-13.38}^{+12.44}$ &
$0.70_{-0.22}^{+0.25}$ &
$0$ &
$0.51_{-0.44}^{+0.42}$ &
$1$ &
$0.21_{-0.98}^{+0.70}$ &
$0.03_{-0.18}^{+0.24}$ &
$0.48$ &
/ \\

GW190521 & $\chi_2=0$ &
$84.18_{-12.17}^{+14.48}$ &
$62.14_{-11.89}^{+10.90}$ &
$0.74_{-0.21}^{+0.22}$ &
$0.45_{-0.37}^{+0.40}$ &
$0$ &
$0.41_{-0.85}^{+0.52}$ &
$1$ &
$0.08_{-0.19}^{+0.24}$ &
/ &
$0.39$ \\

& LVK &
$88.55_{-14.48}^{+15.78}$ &
$60.03_{-13.01}^{+12.82}$ &
$0.68_{-0.20}^{+0.26}$ &
$0.46_{-0.40}^{+0.45}$ &
$0.51_{-0.44}^{+0.41}$ &
$-0.06_{-0.80}^{+0.89}$ &
$0.20_{-0.97}^{+0.70}$ &
$0.02_{-0.41}^{+0.31}$ &
/ &
/ \\

\hline

& $\chi_1=0$ &
$26.51_{-7.32}^{+5.15}$ &
$9.06_{-1.56}^{+5.54}$ &
$0.34_{-0.10}^{+0.39}$ &
$0$ &
$0.52_{-0.40}^{+0.39}$ &
$1$ &
$-0.21_{-0.63}^{+0.87}$ &
$-0.02_{-0.14}^{+0.10}$ &
$-0.58$ &
/ \\

GW240507\_041632 & $\chi_2=0$ &
$29.21_{-5.95}^{+6.00}$ &
$8.67_{-1.33}^{+1.97}$ &
$0.30_{-0.09}^{+0.15}$ &
$0.40_{-0.26}^{+0.34}$ &
$0$ &
$0.00_{-0.39}^{+0.43}$ &
$1$ &
$0.00_{-0.14}^{+0.10}$ &
/ &
$-1.40$ \\

& LVK &
$23.94_{-5.76}^{+7.34}$ &
$10.19_{-2.42}^{+4.05}$ &
$0.43_{-0.19}^{+0.34}$ &
$0.43_{-0.29}^{+0.31}$ &
$0.60_{-0.46}^{+0.33}$ &
$-0.27_{-0.48}^{+0.65}$ &
$-0.02_{-0.62}^{+0.62}$ &
$-0.09_{-0.18}^{+0.17}$ &
/ &
/ \\

\hline

& $\chi_1=0$ &
$33.72_{-0.75}^{+1.05}$ &
$32.52_{-1.16}^{+0.74}$ &
$0.97_{-0.06}^{+0.03}$ &
$0$ &
$0.09_{-0.08}^{+0.22}$ &
$1$ &
$-0.26_{-0.47}^{+0.58}$ &
$-0.01_{-0.04}^{+0.02}$ &
$-0.62$ &
/ \\

GW250114\_082203 & $\chi_2=0$ &
$33.76_{-0.73}^{+1.12}$ &
$32.59_{-1.24}^{+0.74}$ &
$0.97_{-0.06}^{+0.03}$ &
$0.06_{-0.05}^{+0.17}$ &
$0$ &
$-0.26_{-0.59}^{+0.86}$ &
$1$ &
$-0.01_{-0.03}^{+0.02}$ &
/ &
$-0.24$ \\

& LVK &
$33.71_{-0.73}^{+1.07}$ &
$32.54_{-1.27}^{+0.74}$ &
$0.97_{-0.06}^{+0.03}$ &
$0.08_{-0.07}^{+0.17}$ &
$0.09_{-0.08}^{+0.20}$ &
$-0.17_{-0.64}^{+0.92}$ &
$-0.18_{-0.64}^{+0.89}$ &
$-0.02_{-0.03}^{+0.03}$ &
/ &
/ \\

\hline
\end{tabular}%
}
\end{table*}

\subsection{Results}

In this section, we present the impact of spin-prior assumptions on parameter estimation by comparing the posterior distributions obtained under different prior configurations across the selected events. The intrinsic parameters examined in this study include the individual component masses $m_i$, mass ratio $q$, individual spin magnitudes $\chi_i$, spin tilt angles $\theta_i$, and effective inspiral spin $\chi_{\rm eff}$.

\begin{figure*}
    \centering
\includegraphics[width=0.32\textwidth]{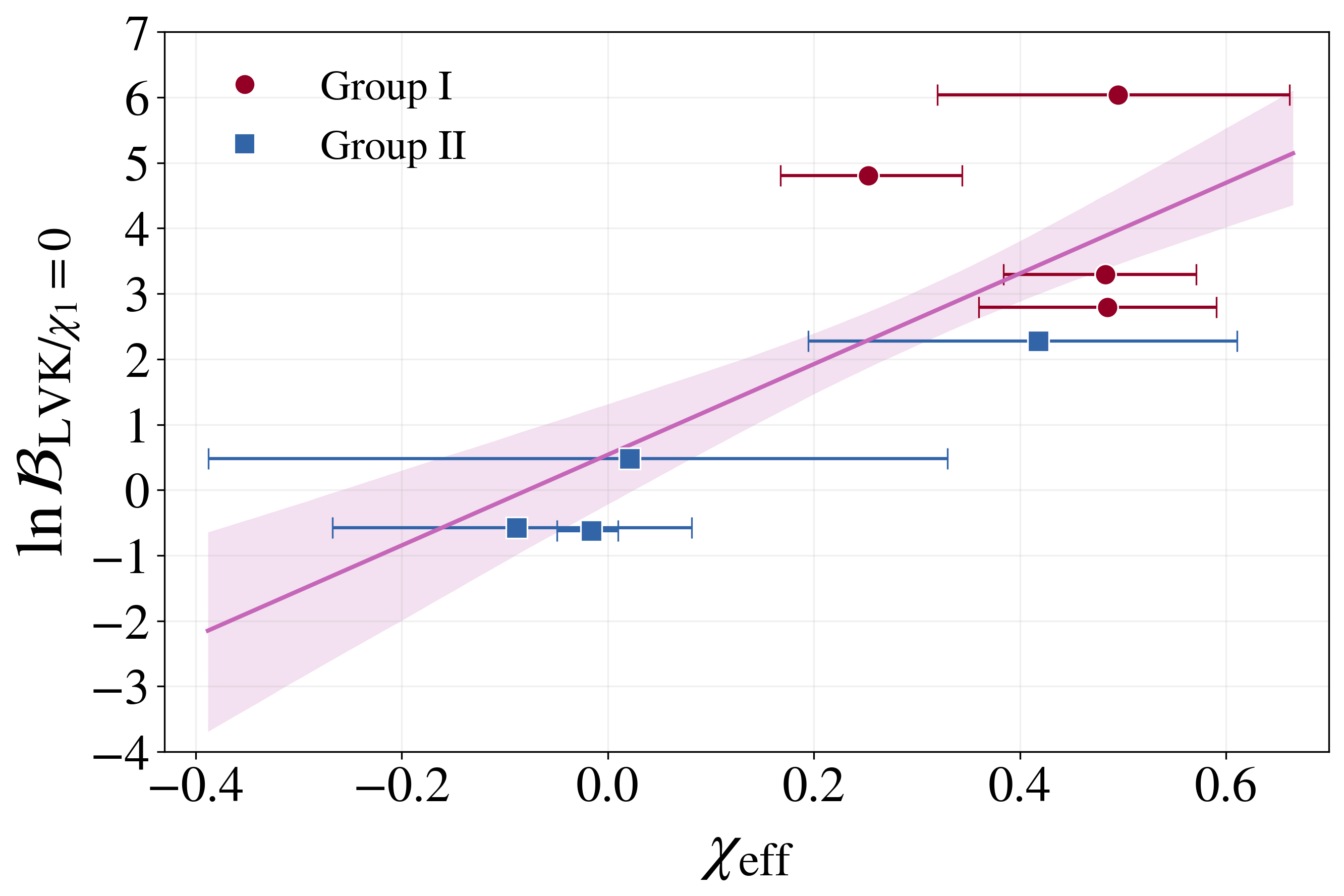}
\includegraphics[width=0.32\textwidth]{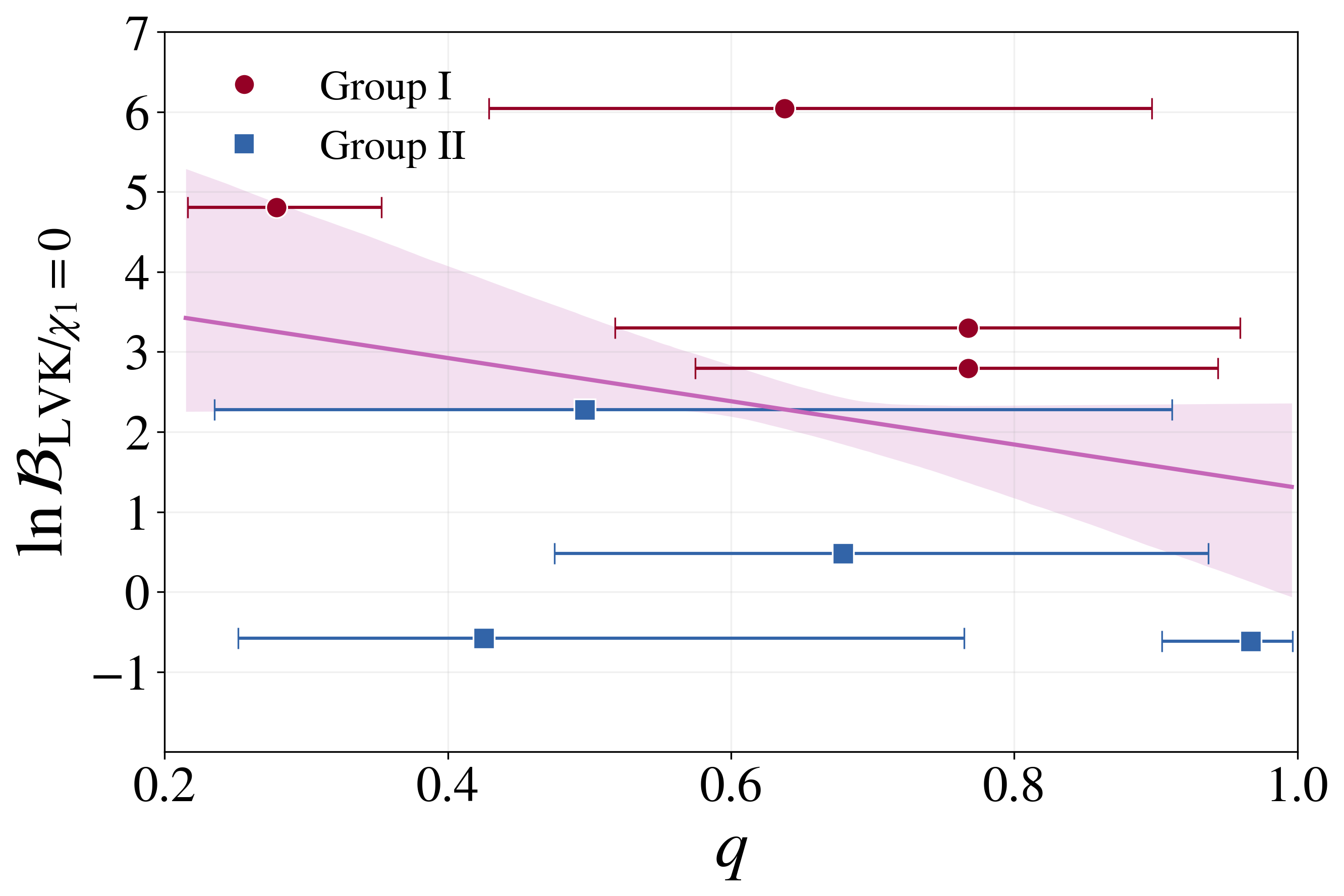}
\includegraphics[width=0.32\textwidth]{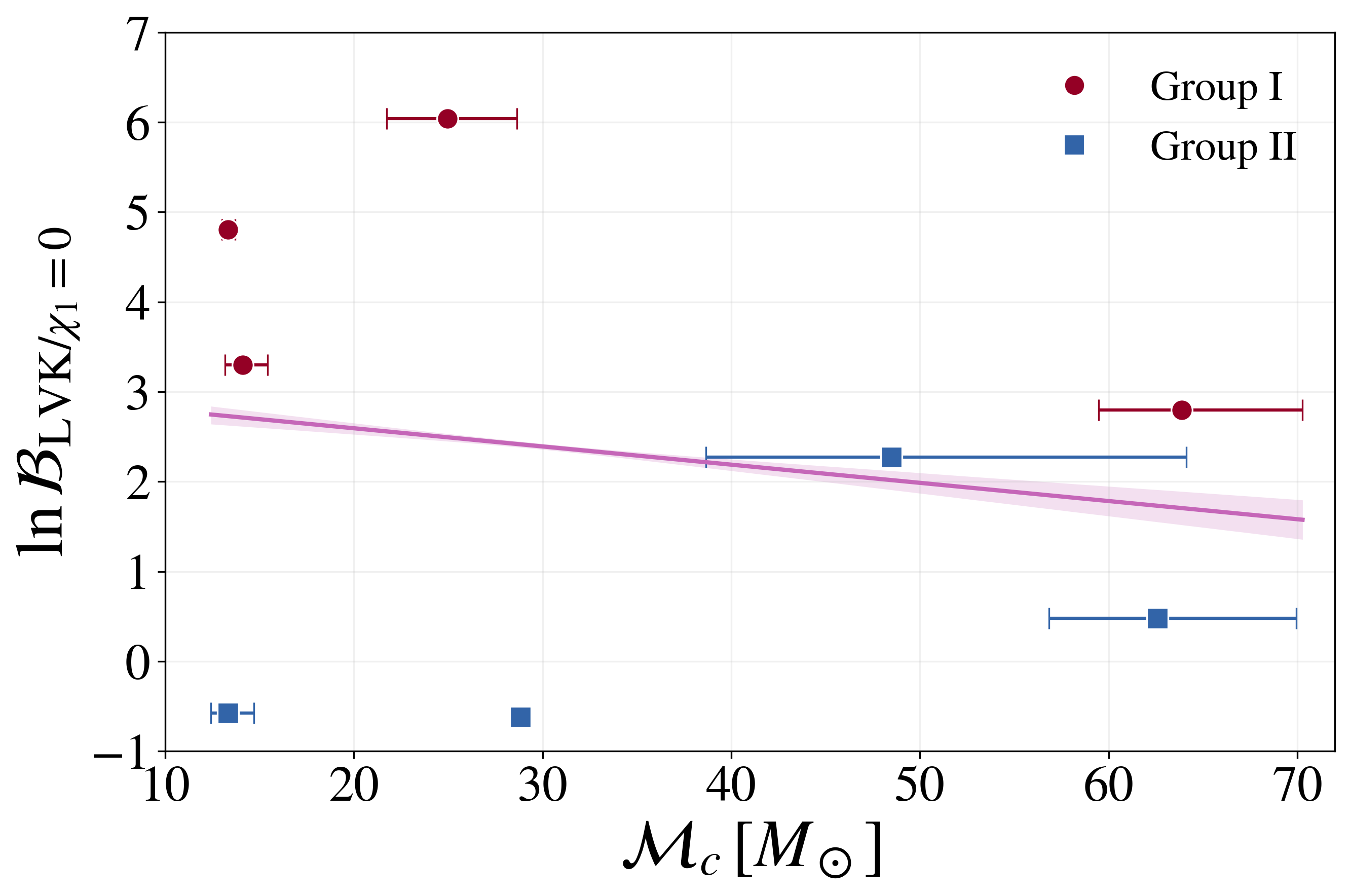}
    \caption{The natural logarithmic Bayes factor, \(\ln \mathcal{B}_{\mathrm{LVK}/\chi_1=0}\), versus the effective inspiral spin $\chi_{\rm eff}$ (left panel), mass ratio $q$ (middle panel), and chirp mass $\mathcal{M_\mathrm{c}}$ (right panel),  with the $\mathrm{LVK}$ priors, for the selected BBH events. The markers represent the median values of \(\chi_{\mathrm{eff}}\) and $q$, and the horizontal error bars correspond to the 90\% credible interval. The magenta line represents the best-fitting linear relation to the median values, with the 90\% credible interval indicated by the purple shaded region. In all three panels, events corresponding to the Group I in Table \ref{table1} are displayed in red markers, while Group II events are displayed in blue markers.} 
    \label{chi_eff_vs_lnBF}
\end{figure*}

Table~\ref{table1} presents the resulting Bayes factors and parameter inferences. In general, we consider $\ln\mathcal{B} > 2.5$ to indicate strong evidence in favor of one model over the other, while $\ln\mathcal{B} < -2.5$ indicates strong evidence against it. Based on these Bayes factors, we divide the systems into two groups according to whether they show strong evidence against the $\chi_1=0$ prior. Group I comprises four events (GW190517\_055101, GW190412, GW241113\_163507, and GW231028\_153006). These events show clear evidence against the $\chi_1=0$ prior, suggesting that a non-spinning primary BH is disfavored for these systems. Interestingly, they are characterized by preferentially high effective inspiral spins, as shown in the left panel of Figure~\ref{chi_eff_vs_lnBF}, while exhibiting weak anti-correlations with the mass ratio $q$ and chirp mass $\mathcal{M}_{\mathrm{c}}$ (see the middle and right panels).
For the remaining systems, which constitute Group II, we find no strong evidence in favor of any particular spin prior. In the following, we present the inferred properties of $\chi_{\rm eff}$, $q$, and the component masses ($m_1$ and $m_2$) for these two groups of events.

\subsubsection{Spin parameters}
The parameter \(\chi_{\mathrm{eff}}\), which represents the mass-weighted projection of the component BHs' spins onto the orbital angular-momentum direction, is particularly well constrained by the gravitational-wave signal during the inspiral. It is defined as
\begin{equation}
\chi_{\rm eff}=\frac{m_1\chi_{\rm 1}\cos\theta_1 +m_2\chi_{\rm 2}\cos\theta_2}{m_1+m_2},
\end{equation}
where $m_1$ and $m_2$ are the component masses, $\chi_1$ and $\chi_2$ are their dimensionless spin magnitudes, and $\theta_1$ and $\theta_2$ are the corresponding spin-tilt angles relative to the orbital angular momentum. 

\begin{figure*}
    \centering
\includegraphics[width=0.99\textwidth]{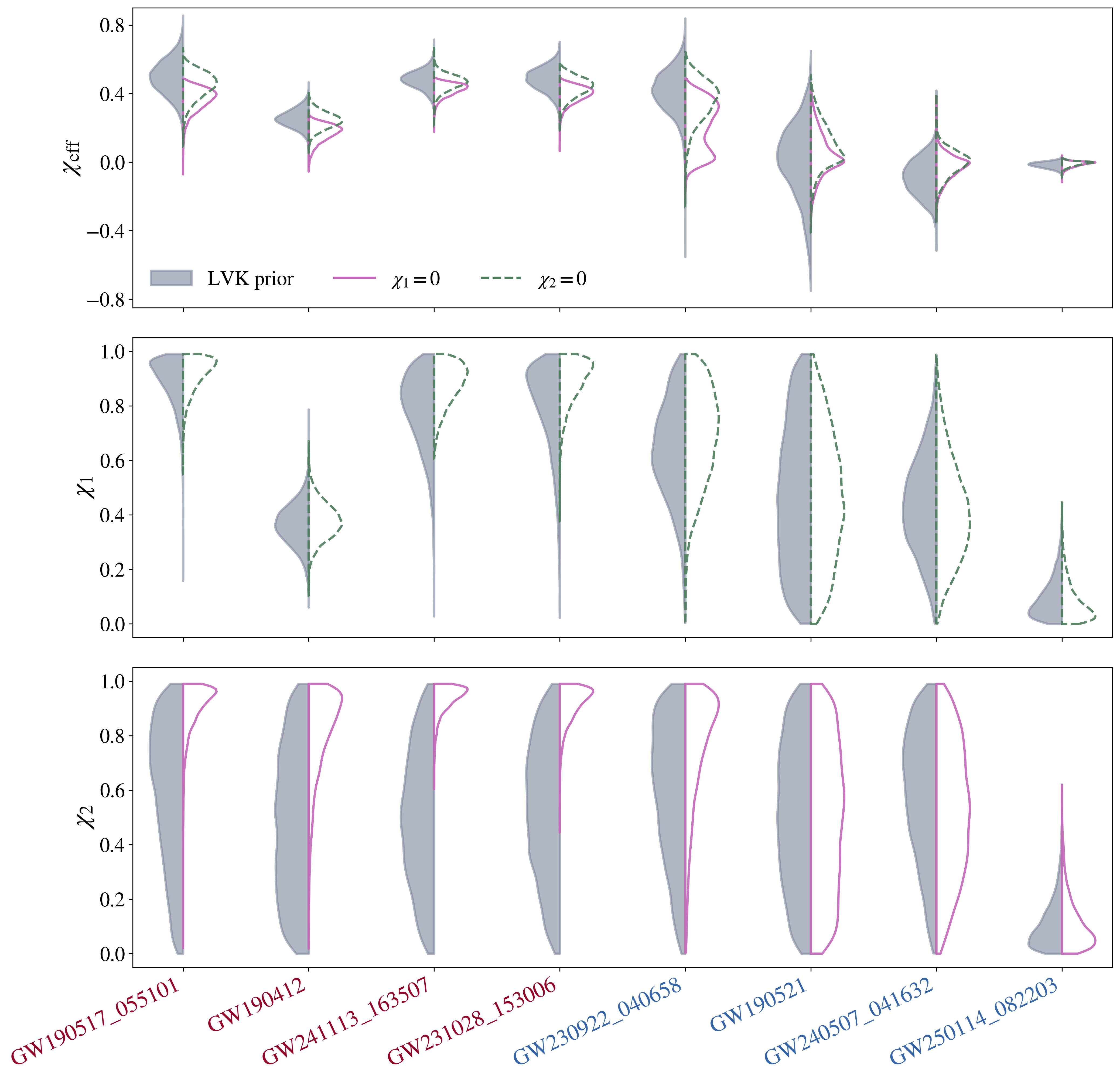}
    \caption{Violin plots showing the marginalized posterior distributions of the effective inspiral spin,  \(\chi_{\rm eff}\), inferred using the LVK default spin prior (shaded region), the \(\chi_1=0\) prior (solid magenta line), and the \(\chi_2=0\) prior (dashed green line). The event names for Group I (II) are colored red (blue).} 
    \label{chi_eff_violins}
\end{figure*}

In the upper panel of Figure~\ref{chi_eff_violins}, we present the inferred posterior distributions of $\chi_{\rm eff}$ under the three spin priors. First, the $\chi_{\rm eff}$ distributions obtained with the $\chi_1 = 0$ prior shift towards lower values in Group I compared with those obtained with the $\mathrm{LVK}$ and $\chi_2 = 0$ priors. As shown in the left panel of Figure~\ref{chi_eff_vs_lnBF}, the events in Group I exhibit a clear preference against the $\chi_1 = 0$ prior. This indicates that the assumption of a non-spinning primary BH, although astrophysically motivated, is less favored by the data. In contrast, the inferred results obtained with the $\mathrm{LVK}$ and $\chi_2 = 0$ priors are both viable, as supported by the corresponding Bayes factors.

To obtain a more precise characterization of the relationship between $\ln\mathcal{B}_{\mathrm{LVK}/\chi_1=0}$ and $\chi_{\rm eff}$, we fit $y_i=\ln\mathcal{B}_{\mathrm{LVK}/\chi_1=0,i}$ as a linear function of $x=\chi_{\rm eff}$ ($i$ refers to individual evnet). Each $y_i$ is treated as fixed, while its conditional
distribution is modeled as a Gaussian,
\[
p(y_i\mid x,m,b,\sigma_{\mathrm{int}})
=
\mathcal{N}\!\left(y_i\mid mx+b,\sigma_{\mathrm{int}}^2\right),
\]
where $m$ is the correlation slope, $b$ intercept, and $\sigma_{\mathrm{int}}$ intrinsic scatter. The uncertainty in $x$ is incorporated by marginalizing this Gaussian
likelihood over the full LVK posterior $p_i(x)$,
\[
\mathcal{L}_i
=
\int p_i(x)\,
\mathcal{N}\!\left(y_i\mid mx+b,\sigma_{\mathrm{int}}^2\right)\,dx,
\]
which is evaluated directly using the posterior samples (i.e., $\chi_{\rm eff}$). Finding the maximized combination of Gaussian likelihoods for the eight events gives $m$, $b$, and
$\sigma_{\mathrm{int}}$, where
$\sigma_{\mathrm{int}}$ quantifies the event-to-event dispersion around the
linear relation.

The linear relation above provides a simple way to estimate how likely the $\chi_1=0$ prior is strongly disfavored, by only using its $\chi_{\rm eff}$ posteriors inferred with the default prior, avoiding to calculate the computational expensive ``Bayes evidence''  for the $\chi_1=0$ prior. Assuming the $\chi_1=0$ prior for the standard common envelope evolution channel \cite[i.e., CEE, ][]{Belczynski2016}, we can further derive the merger rate for BBHs \textit{not} originated from this channel by incorporating the $\ln \mathcal{B}$-$\chi_{\rm eff}$ relation and the ``event-based'' merger rates. The ``event-based'' merger rates are calculated based upon the inferred properties of specific events. This approach treats individual events as distinct classes, and the total event rate is the sum of the individual rates \citep{2016PhRvX...6d1015A}. To estimate the ``non-CEE'' merger rate, we follow the steps below:
\begin{enumerate}[label=\arabic*., leftmargin=*]
  \item \textit{Calculate sensitive space-time volume ${\rm VT}$ for all BBH events with false alarm rate $<1 \rm{yr}^{-1}$.} The population-averaged sensitive space-time volume of the search ($\left \langle {\rm VT} \right \rangle $) is 
  \begin{equation}
      \langle VT \rangle = T \int \mathrm{d}z \mathrm{d}\theta \frac{\mathrm{d}V_c}{\mathrm{d}z} \frac{1}{1+z} s(\theta) f(z, \theta),
  \label{eq:vt}
  \end{equation}
   where $V_c$ is the comoving volume, $\theta$ is the parameters describing the properties (masses and spins) of BBHs, $s(\theta)$ is the distribution function for the astrophysical population, and $f(z, \theta)$ is the selection function giving the probability of detecting a source with parameters $\theta$ at redshift $z$ \citep{2016PhRvX...6d1015A}. Since we are calculating the space-time volume for each event, $s(\theta)$ is treated as the joint posterior distribution of $(m_1,m_2,a_1,a_2,\cos \theta_1, \cos \theta_2)$ inferred with the default prior. The injections released by \citet{gwtc5_pop} is used to calculate E.q.(\ref{eq:vt}) numerically. 
  
  \item \textit{Derive the probability distribution of event-like merger rates for all BBHs.} With the $\left \langle {\rm VT} \right \rangle_i $ (here $i$ represent the $i$-th event) obtained in the above step and consider Poisson fluctuations, the likelihood for event-like (system with masses and spins similar to a specific event) merger rate is:
  \begin{equation}
      \mathcal{L}(\mathcal{R}_i) = \mathrm{Poisson}(\mathcal{R}_i\left \langle VT \right \rangle_i \mid N_{\rm obs}=1 ),
  \label{eq:rate}
  \end{equation}
  where $\mathrm{Poisson}$ is the Poisson distribution. Adopting an uniform prior ($\pi(\mathcal{R}_i)$) ranging from  $\frac{1}{100\left \langle VT \right \rangle_i}$ to $\frac{10}{\left \langle VT \right \rangle_i}$, the posterior distribution can be derived by $\mathcal{R}_i \propto \mathcal{L}(\mathcal{R}_i)\pi(\mathcal{R}_i)$.
  
  \item \textit{Obtain the lower limit for non-CEE merger rate using Monte Carlo approach.} Based on the $\ln \mathcal{B}$-$\chi_{\rm eff}$ relation, the probability that the $\chi_1=0$ prior is strongly disfavored for a specific posterior sample $\chi_{{\rm eff}}$ is $\mathrm{P} = \mathrm{CDF}_\mathcal{N}(\ln \mathcal{B}=2.5 \mid \chi_{\rm eff})-1$. For the $i$-th event, we randomly draw a $\chi_{{\rm eff},i}$ sample and a $\mathcal{R}_i$ sample from their posterior distributions, then the lower limit for non-CEE merger rate is calculated by $\mathcal{R}_{{\rm tot}} = \sum_{i}\mathrm{P}_i\mathcal{R}_i$. We repeat this procedure 30000 times to obtain $\mathcal{R}_{{\rm tot}}$ samples whose statistical properties reflect the uncertainties of $\chi_{{\rm eff},i}$ and $\mathcal{R}_i$. Finally, we derive $\mathcal{R}_{{\rm tot}} = 12.2^{+2.9}_{-2.2}\ {\rm Gpc}^{-3}{\rm yr}^{-1}$.
\end{enumerate}

Interestingly, the $\chi_2 = 0$ prior can be motivated by the mass-ratio reversal scenario. This scenario has been investigated in recent population synthesis studies \cite[e.g.,][]{Broekgaarden2022,Hu2026}. In this scenario, the initially less massive star accretes mass from its initially more massive companion and eventually becomes the more massive component. If the initially less massive star subsequently forms the second-born BH, this BH may be expected to have a negligible natal spin, while the first-born BH becomes the less massive component. Notably, under the $\chi_1 = 0$ prior, the inferred spin distributions of the primary BHs are shifted towards higher values and are more tightly concentrated than those obtained with the $\mathrm{LVK}$ prior, particularly for the events in Group I (see the bottom panel).

For the events in Group II, the inferred results under all three spin priors are viable, as there is no statistically significant evidence favoring or disfavoring any one prior over the others. Moreover, their inferred parameters exhibit only small deviations from those of the events in Group I.

\subsubsection{Mass ratio}
\begin{figure*}
    \centering
\includegraphics[width=0.99\textwidth]{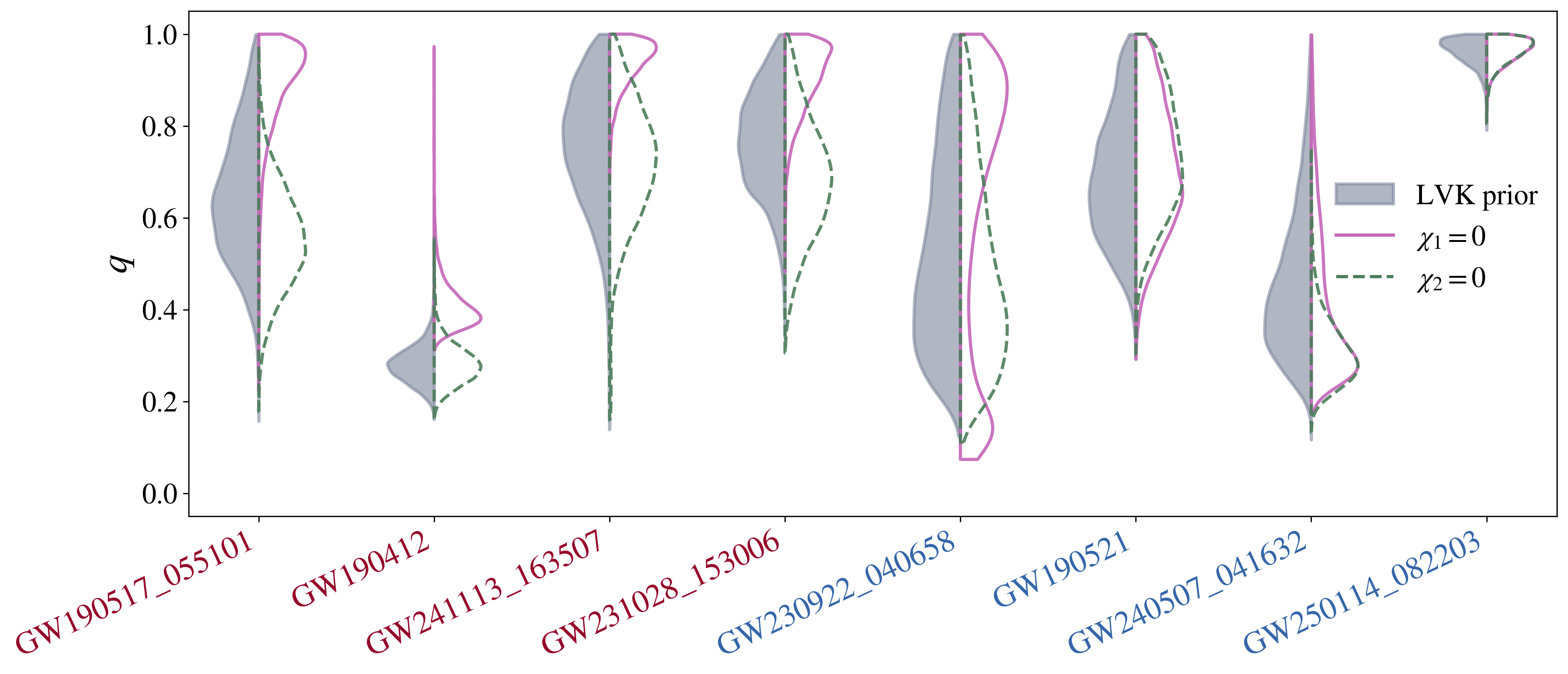}
    \caption{As in Fig.~\ref{chi_eff_violins}, but for the mass ratio \(q\).} 
    \label{q_violins}
\end{figure*}

In Fig.~\ref{q_violins}, we present the posterior distributions of the mass ratio under the three spin priors. As discussed above, the $\chi_1=0$ prior is less favored than the other priors for events in Group I. Statistically, the results obtained with the $\chi_2=0$ prior are comparable to those obtained with the $\mathrm{LVK}$ prior, although the corresponding mass-ratio distributions tend to shift towards lower values. This is more apparent when compared with the inferred $\chi_{\rm eff}$ distributions (see Fig.~\ref{chi_eff_violins}).

In contrast, the events in Group II yield statistically indistinguishable results under the three spin priors, with similar mass-ratio distributions across the different priors. In particular, GW230922\_040658 exhibits a bimodal posterior distribution, with a structure similar to that shown in the $\chi_{\rm eff}$ posterior distribution.

\subsubsection{Component masses}
The chirp mass, $\mathcal{M_\mathrm{c}}$, is defined as
\begin{equation}
\mathcal{M_\mathrm{c}} = \frac{\left( m_1 m_2 \right)^{3/5}}{\left( m_1 + m_2 \right)^{1/5}},
\end{equation}\label{chirp_mass}

\noindent
where $m_1$ and $m_2$ denote the component masses of the primary and secondary BHs, respectively.

Using the above equation, we derive the individual component masses as
\begin{equation}
    m_1 = \mathcal{M}_{\rm c}(1+q)^{\frac{1}{5}}q^{-\frac{3}{5}},
    \quad
    m_2 = \mathcal{M}_{\rm c}(1+q)^{\frac{1}{5}}q^{\frac{2}{5}}.
\end{equation}

At fixed chirp mass, their sensitivities to the mass ratio can be expressed as
\begin{equation}
    \frac{\mathrm{d}\ln m_1}{\mathrm{d}\ln q}
    = -\frac{3+2q}{5(1+q)},
    \quad
    \frac{\mathrm{d}\ln m_2}{\mathrm{d}\ln q}
    = \frac{2+3q}{5(1+q)}.
\end{equation}
Thus, for $q<1$, we have
\begin{equation}
    \left|\frac{\mathrm{d}\ln m_1}{\mathrm{d}\ln q}\right|
    >
    \left|\frac{\mathrm{d}\ln m_2}{\mathrm{d}\ln q}\right|.
\end{equation}

Therefore, for $q<1$ ($q = m_2/m_1$), the primary mass $m_1$ exhibits a stronger sensitivity to variations in the mass ratio $q$ than the secondary mass $m_2$.

\begin{figure*}
    \centering
    \includegraphics[width=0.99\textwidth]{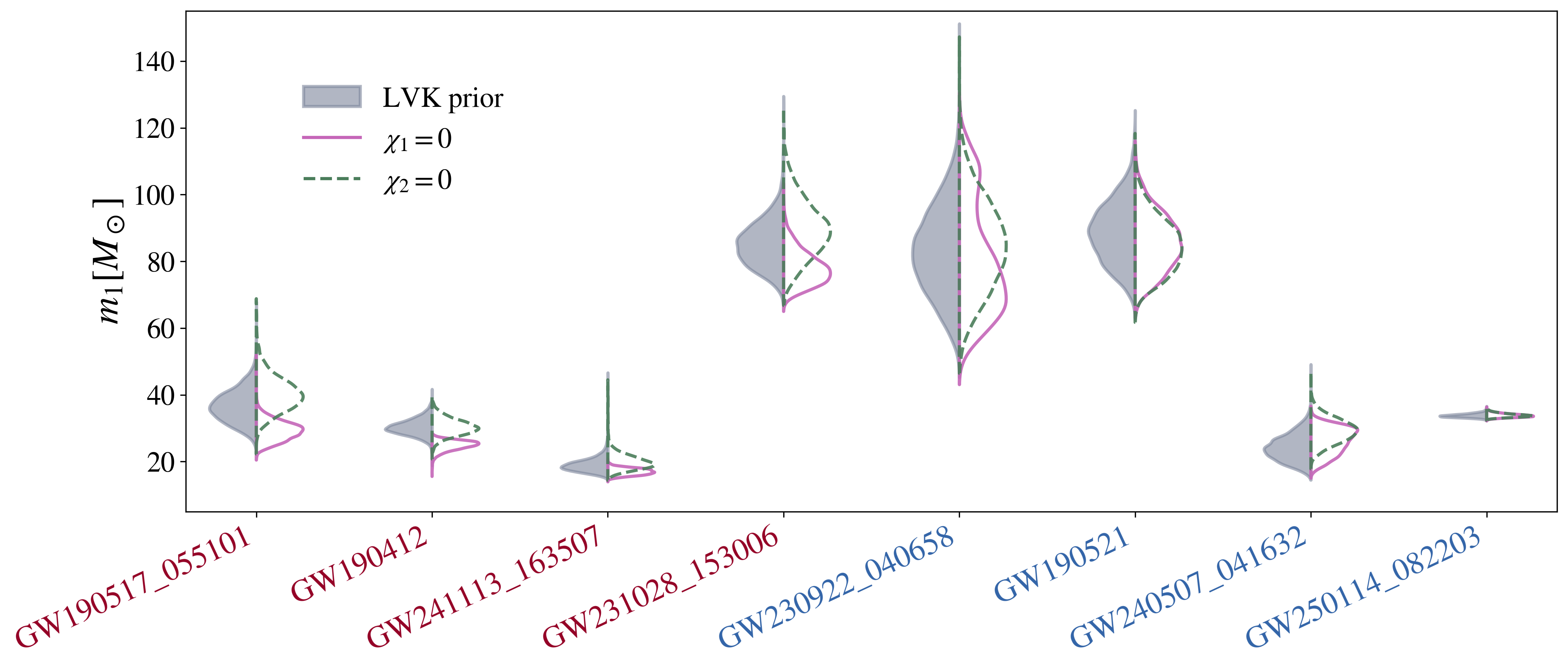}
    \caption{As in Figure~\ref{chi_eff_violins}, but for the primary mass \(m_1\).} 
    \label{m1_violins}
\end{figure*}

\begin{figure*}
    \centering
    \includegraphics[width=0.99\textwidth]{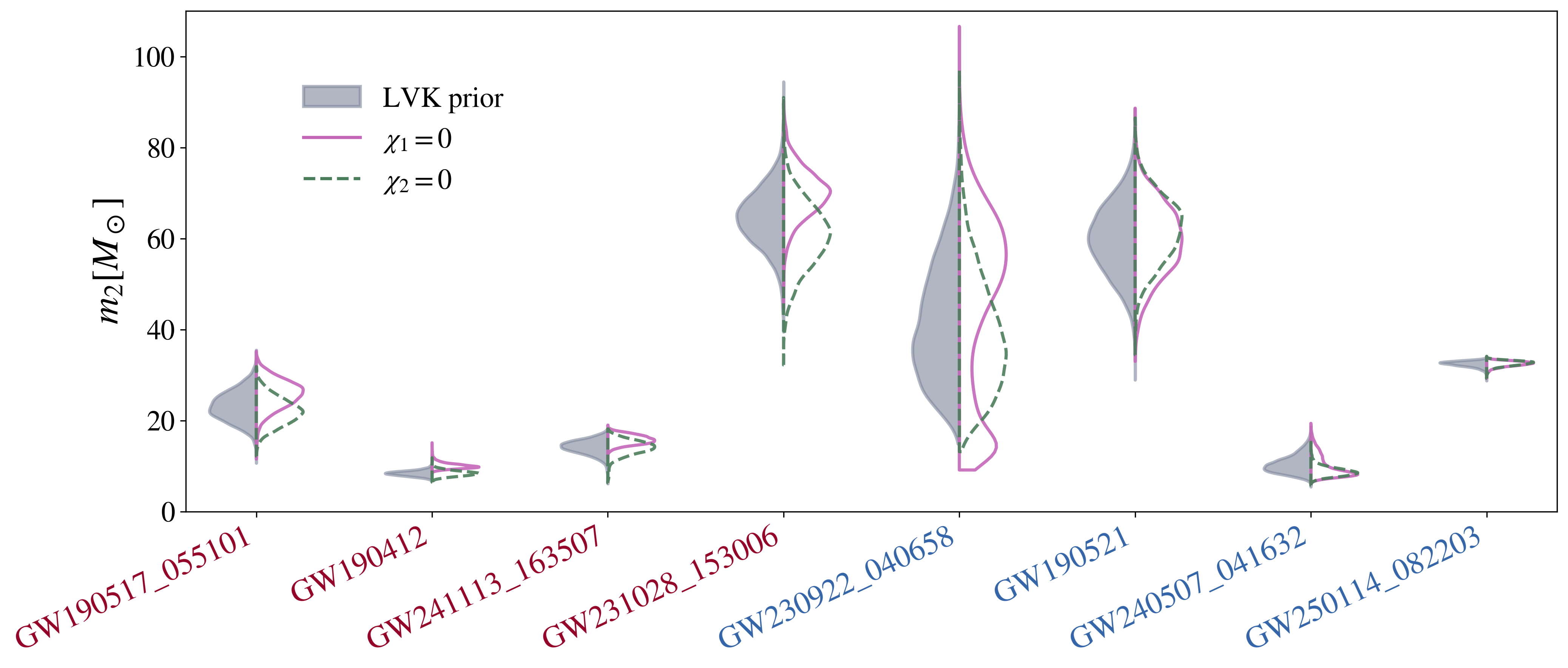}
    \caption{As in Figure~\ref{chi_eff_violins}, but for the secondary mass \(m_2\).} 
    \label{m2_violins}
\end{figure*}

In Figures \ref{m1_violins} and \ref{m2_violins}, we show the marginalized posterior distributions of the two component masses for all events. In general, the inferred chirp mass is much less sensitive to the choice of spin prior than parameters such as $\chi_{\rm eff}$ and the mass ratio $q$. Consequently, variations in the inferred mass ratio redistribute the total mass between $m_1$ and $m_2$, while the chirp mass remains relatively well constrained. In particular, a decrease in $m_1$ is accompanied by an increase in $m_2$ at fixed chirp mass \cite[see also][]{Xue2025}. This behavior is clearly evident for the events in Group I under the different spin priors, particularly for those exhibiting substantial shifts in $q$. For instance, GW231028\_153006 shows a shift from $m_1=84.9\,M_\odot$ and $m_2=64.7\,M_\odot$ under the $\mathrm{LVK}$ prior to $m_1=89.2\,M_\odot$ and $m_2=60.7\,M_\odot$ under the $\chi_1=0$ prior.

In addition to the inferred properties, including $\chi_{\rm eff}$, $q$, and the component masses ($m_1$ and $m_2$), the spin tilt angles ($\theta_1$ and $\theta_2$) can also vary under different spin priors, which are not explored in the present work. Misalignment between the BH spin and the orbital angular momentum remains possible in merging BBHs formed through the isolated binary evolution.

\section{Conclusions and Discussion}\label{sect4}
In this work, we perform Bayesian inference for a subset of eight BBH events spanning a wide range of properties in the $\mathcal{M}_{\rm c}$-$q$-$\chi_{\rm eff}$ parameter space. To investigate the impact of spin priors on the inferred parameters, we consider three different priors: the $\mathrm{LVK}$, $\chi_1 = 0$, and $\chi_2 = 0$. The $\mathrm{LVK}$ prior, which is adopted by default in the $\mathrm{LVK}$ data analysis, is generally considered to be noninformative. In contrast, the $\chi_1 = 0$ and $\chi_2 = 0$ priors are astrophysically motivated by binary massive-star evolution, in which one of the BH is expected to have negligible spin.

We first identify a positive correlation between the Bayes factor, $\ln \mathcal{B}_{\mathrm{LVK}/\chi_1=0}$, and the effective inspiral spin $\chi_{\rm eff}$, along with very weak anti-correlations with the mass ratio $q$ and chirp mass $\mathcal{M}_{\rm c}$. Applying the first relationship, we obtain the total merger‑rate density  $\mathcal{R}_{{\rm tot}} = 12.2^{+2.9}_{-2.2}\ {\rm Gpc}^{-3}{\rm yr}^{-1}$ for BBHs formed via channels other than the CEE channel (i.e., non-CEE channel). This result suggests that the assumption of a non-spinning primary BH is disfavored for the Group I events (GW190517\_055101, GW190412, GW241113\_163507, and GW231028\_153006), which have relatively high $\chi_{\rm eff}$. For these four events, however, the inferred results are indistinguishable under the $\mathrm{LVK}$ and $\chi_2=0$ priors, indicating that the $\chi_2=0$ (i.e., MRR scenario) prior may provide a plausible interpretation of their formation history. 

In this scenario, the progenitor binary undergoes mass-ratio reversal during the first stable mass-transfer phase through Roche-lobe overflow. The initially more massive star loses most of its hydrogen-rich envelope, becomes a helium star, and eventually forms the first-born BH (the primary). As expected, this BH is expected to be born with negligible spin \citep{Qin2018,fuller2019}. Meanwhile, the initially less massive star accretes sufficient mass during the first mass-transfer phase, resulting in mass-ratio reversal and ultimately forming the more massive, second-born BH. Its immediate progenitor, a helium star in a close binary, can be efficiently spun up through tidal interactions, leading to the formation of a rapidly spinning BH \citep{Detmers2008,Qin2018,Zaldarriaga2018,Bavera2020,Hu2022,Ma2023,Wang2026}. This mass-ratio-reversal scenario has also been proposed as an alternative formation channel for GW241011\_233834 \citep{Hu2026}.

For the remaining events in Group II, their inferred results under all three spin priors are viable, as there is no statistically significant evidence favoring or disfavoring any one prior over the others. Moreover, their inferred parameters exhibit relatively small deviations from those of the events in Group I.

As discussed above, GW190517\_055101 could originate from the standard isolated binary evolution channel involving the mass ratio reversal. In this scenario, the newly inferred primary spin is remarkably high, $\chi_1 = 0.924_{-0.196}^{+0.061}$. However, \cite{Qin2022} suggested that this event could instead be formed through the chemically homogeneous evolution channel based on the combined priors predicted by \cite{Zevin2021}. The origin of GW190412 was first investigated by \cite{Mandel2020}, who argued that the system could be alternatively explained by a highly spinning secondary and a nearly non-spinning primary, which is astrophysically well motivated. Subsequently, \cite{Zevin2020} found strong evidence against the non-spinning primary hypothesis in the data, which is consistent with our results. We also highlight that the non‑spinning primary BH can also be produced via stable mass‑transfer \cite[e.g.,][]{van2017,Bavera2021,Olejak2021}, in contrast to the canonical common‑envelope channel. In general, BHs can acquire natal kicks at birth, which can misalign BH‑spin axes relative to the orbital angular momentum \citep[e.g.,][]{Kalogera2000}. Such kicks may originate from asymmetric mass ejection \citep[e.g.,][]{Janka2013} or anisotropic neutrino emission during core collapse \citep[e.g.,][]{Fryer2006}. As an alternative mechanism, the spin axes of newly formed BHs can experience intrinsic reorientation via spin‑axis tossing \citep{Tauris2022}.

\section*{Acknowledgements}
Wei-Hua Guo acknowledges support from the Scientific Research Startup Foundation of Nanyang Normal University. This research has made use of data and software obtained from the Gravitational Wave Open Science Center (https://www.gw-openscience.org), a service of LIGO Laboratory, the LIGO Scientific Collaboration and the Virgo Collaboration.


\bibliographystyle{mnras}
\bibliography{ref} 

@ARTICLE{2016PhRvX...6d1015A,
       author = {{Abbott}, B.~P. and {Abbott}, R. and {Abbott}, T.~D. and {Abernathy}, M.~R. and {Acernese}, F. and {Ackley}, K. and {Adams}, C. and {Adams}, T. and {Addesso}, P. and {Adhikari}, R.~X. and {Adya}, V.~B. and {Affeldt}, C. and {Agathos}, M. and {Agatsuma}, K. and {Aggarwal}, N. and {Aguiar}, O.~D. and {Aiello}, L. and {Ain}, A. and {Ajith}, P. and {Allen}, B. and {Allocca}, A. and {Altin}, P.~A. and {Anderson}, S.~B. and {Anderson}, W.~G. and {Arai}, K. and {Araya}, M.~C. and {Arceneaux}, C.~C. and {Areeda}, J.~S. and {Arnaud}, N. and {Arun}, K.~G. and {Ascenzi}, S. and {Ashton}, G. and {Ast}, M. and {Aston}, S.~M. and {Astone}, P. and {Aufmuth}, P. and {Aulbert}, C. and {Babak}, S. and {Bacon}, P. and {Bader}, M.~K.~M. and {Baker}, P.~T. and {Baldaccini}, F. and {Ballardin}, G. and {Ballmer}, S.~W. and {Barayoga}, J.~C. and {Barclay}, S.~E. and {Barish}, B.~C. and {Barker}, D. and {Barone}, F. and {Barr}, B. and {Barsotti}, L. and {Barsuglia}, M. and {Barta}, D. and {Bartlett}, J. and {Bartos}, I. and {Bassiri}, R. and {Basti}, A. and {Batch}, J.~C. and {Baune}, C. and {Bavigadda}, V. and {Bazzan}, M. and {Bejger}, M. and {Bell}, A.~S. and {Berger}, B.~K. and {Bergmann}, G. and {Berry}, C.~P.~L. and {Bersanetti}, D. and {Bertolini}, A. and {Betzwieser}, J. and {Bhagwat}, S. and {Bhandare}, R. and {Bilenko}, I.~A. and {Billingsley}, G. and {Birch}, J. and {Birney}, R. and {Birnholtz}, O. and {Biscans}, S. and {Bisht}, A. and {Bitossi}, M. and {Biwer}, C. and {Bizouard}, M.~A. and {Blackburn}, J.~K. and {Blair}, C.~D. and {Blair}, D.~G. and {Blair}, R.~M. and {Bloemen}, S. and {Bock}, O. and {Boer}, M. and {Bogaert}, G. and {Bogan}, C. and {Bohe}, A. and {Bond}, C. and {Bondu}, F. and {Bonnand}, R. and {Boom}, B.~A. and {Bork}, R. and {Boschi}, V. and {Bose}, S. and {Bouffanais}, Y. and {Bozzi}, A. and {Bradaschia}, C. and {Brady}, P.~R. and {Braginsky}, V.~B. and {Branchesi}, M. and {Brau}, J.~E. and {Briant}, T. and {Brillet}, A. and {Brinkmann}, M. and {Brisson}, V. and {Brockill}, P. and {Broida}, J.~E. and {Brooks}, A.~F. and {Brown}, D.~A. and {Brown}, D.~D. and {Brown}, N.~M. and {Brunett}, S. and {Buchanan}, C.~C. and {Buikema}, A. and {Bulik}, T. and {Bulten}, H.~J. and {Buonanno}, A. and {Buskulic}, D. and {Buy}, C. and {Byer}, R.~L. and {Cabero}, M. and {Cadonati}, L. and {Cagnoli}, G. and {Cahillane}, C. and {Calder{\'o}n Bustillo}, J. and {Callister}, T. and {Calloni}, E. and {Camp}, J.~B. and {Cannon}, K.~C. and {Cao}, J. and {Capano}, C.~D. and {Capocasa}, E. and {Carbognani}, F. and {Caride}, S. and {Casanueva Diaz}, J. and {Casentini}, C. and {Caudill}, S. and {Cavagli{\`a}}, M. and {Cavalier}, F. and {Cavalieri}, R. and {Cella}, G. and {Cepeda}, C.~B. and {Cerboni Baiardi}, L. and {Cerretani}, G. and {Cesarini}, E. and {Chamberlin}, S.~J. and {Chan}, M. and {Chao}, S. and {Charlton}, P. and {Chassande-Mottin}, E. and {Cheeseboro}, B.~D. and {Chen}, H.~Y. and {Chen}, Y. and {Cheng}, C. and {Chincarini}, A. and {Chiummo}, A. and {Cho}, H.~S. and {Cho}, M. and {Chow}, J.~H. and {Christensen}, N. and {Chu}, Q. and {Chua}, S. and {Chung}, S. and {Ciani}, G. and {Clara}, F. and {Clark}, J.~A. and {Cleva}, F. and {Coccia}, E. and {Cohadon}, P.-F. and {Colla}, A. and {Collette}, C.~G. and {Cominsky}, L. and {Constancio}, M. and {Conte}, A. and {Conti}, L. and {Cook}, D. and {Corbitt}, T.~R. and {Cornish}, N. and {Corsi}, A. and {Cortese}, S. and {Costa}, C.~A. and {Coughlin}, M.~W. and {Coughlin}, S.~B. and {Coulon}, J.-P. and {Countryman}, S.~T. and {Couvares}, P. and {Cowan}, E.~E. and {Coward}, D.~M. and {Cowart}, M.~J. and {Coyne}, D.~C. and {Coyne}, R. and {Craig}, K. and {Creighton}, J.~D.~E. and {Cripe}, J. and {Crowder}, S.~G. and {Cumming}, A.},
        title = "{Binary Black Hole Mergers in the First Advanced LIGO Observing Run}",
      journal = {Physical Review X},
         year = 2016,
        month = oct,
       volume = {6},
       number = {4},
          eid = {041015},
        pages = {041015},
          doi = {10.1103/PhysRevX.6.041015},
archivePrefix = {arXiv},
       eprint = {1606.04856},
 primaryClass = {gr-qc},
       adsurl = {https://ui.adsabs.harvard.edu/abs/2016PhRvX...6d1015A}
}

@ARTICLE{gwtc5_pop,
       author = {{The LIGO Scientific Collaboration} and {the Virgo Collaboration} and {the KAGRA Collaboration} and {Abac}, A.~G. and {Abe}, A. and {Abouelfettouh}, I. and {Acernese}, F. and {Ackley}, K. and {Adam}, A. and {Adhicary}, S. and {Adhikari}, D. and {Adhikari}, R.~X. and {Adkins}, V.~K. and {Afroz}, S. and {Agapito}, A. and {Agarwal}, D. and {Agathos}, M. and {Aggarwal}, N. and {Aggarwal}, S. and {Aguiar}, O.~D. and {Ahrend}, I.-L. and {Aiello}, L. and {Ain}, A. and {Ajith}, P. and {Akutsu}, T. and {Albers}, L. and {Ali}, W. and {Al-Kershi}, S. and {Allene}, C. and {Allocca}, A. and {Al-Shammari}, S. and {Alvarez}, J.~A. and {Alvarez-Lopez}, S. and {Amar}, W. and {Amarasinghe}, O. and {Amato}, A. and {Amicucci}, F. and {Amra}, C. and {Anand}, A.~B. and {Anand}, C. and {Ananyeva}, A. and {Anderson}, S.~B. and {Anderson}, W.~G. and {Andia}, M. and {Ando}, M. and {Andrade-Oliveira}, F. and {Andr{\'e}s-Carcasona}, M. and {Andrey}, J.~L. and {Andri{\'c}}, T. and {Anglin}, J. and {Anna}, J. and {Antelis}, J.~M. and {Antier}, S. and {Antonini}, F. and {Aoki}, T. and {Aoumi}, M. and {Appavuravther}, E.~Z. and {Appelt}, E.~A. and {Appert}, S. and {Apple}, S.~K. and {Arai}, K. and {Araya}, A. and {Araya}, M.~C. and {Arca Sedda}, M. and {Arciprete}, F. and {Areeda}, J.~S. and {Aritomi}, N. and {Armato}, F. and {Armstrong}, S. and {Arnaud}, N. and {Arogeti}, M. and {Aronson}, S.~M. and {Ashton}, G. and {Aso}, Y. and {Asprea}, L. and {Assiduo}, M. and {Assis de Souza Melo}, S. and {Aston}, S.~M. and {Astone}, P. and {Aswathi}, P.~S. and {Attadio}, F. and {Aubin}, F. and {AultONeal}, K. and {Avallone}, G. and {Avdeev}, N. and {Avila}, E.~A. and {Babak}, S. and {Badger}, C. and {Bae}, S. and {Bagnasco}, S. and {Baimukhametova}, S. and {Baiotti}, L. and {Baka}, T. and {Baker}, K.~A. and {Baker}, T. and {Balbi}, G. and {Baldi}, G. and {Baldicchi}, N. and {Ball}, M. and {Ballardin}, G. and {Ballelli}, M. and {Ballmer}, S.~W. and {Banagiri}, S. and {Banerjee}, B. and {Bankar}, D. and {Baptiste}, T.~M. and {Baral}, P. and {Baratti}, M. and {Barayoga}, J.~C. and {Baric}, K. and {Barish}, B.~C. and {Barker}, D. and {Barman}, N. and {Barone}, F. and {Barr}, B. and {Barrios}, M. and {Barsotti}, L. and {Barsuglia}, M. and {Barta}, D. and {Barton}, M.~A. and {Bartos}, I. and {Basalaev}, A. and {Bassiri}, R. and {Basti}, A. and {Bawaj}, M. and {Bayley}, J.~C. and {Baylor}, A.~C. and {Baynard}, II, P.~A. and {Bazzan}, M. and {Bedakihale}, V.~M. and {Beirnaert}, F. and {Bejger}, M. and {Bell}, A.~S. and {Bellani}, C. and {Bellie}, D.~S. and {Beltran-Martinez}, D. and {Benedetti}, E. and {Benoit}, W. and {Bentara}, I. and {Ben Yaala}, M. and {Bera}, S. and {Bergamin}, F. and {Berger}, B.~K. and {Beroiz}, M. and {Berry}, C.~P.~L. and {Berry}, I. and {Bersanetti}, D. and {Bertheas}, T. and {Bertolini}, A. and {Betzwieser}, J. and {Beveridge}, D. and {Bevins}, N. and {Bezerra-Sobrinho}, J. and {Bhandare}, R. and {Bhatt}, R. and {Bhattacharjee}, A. and {Bhattacharjee}, D. and {Bhattacharyya}, S. and {Bhaumik}, S. and {Biancalana}, V. and {Bianchi}, F. and {Bilenko}, I.~A. and {Bilicki}, M. and {Billingsley}, G. and {Binetti}, A. and {Bini}, S. and {Biot}, S. and {Birnholtz}, O. and {Biscoveanu}, S. and {Bisht}, A. and {Bitossi}, M. and {Bizouard}, M.-A. and {Blaber}, S. and {Blackburn}, J.~K. and {Blagg}, L.~A. and {Blair}, C.~D. and {Blair}, D.~G. and {Bloch}, M. and {Bode}, N. and {Boettner}, N. and {Bogdan}, P. and {Boileau}, G. and {Boldrini}, M. and {Bolingbroke}, G.~N. and {Bonavena}, L.~D. and {Bonhomme}, V.~A. and {Bonilla}, E. and {Bonilla}, M.~S. and {Bonino}, A. and {Bonnand}, R. and {Borchers}, A. and {Borghi}, N. and {Boschi}, V. and {Bose}, S. and {Bossilkov}, V. and {Bothra}, Y. and {Boudon}, A. and {Boybeyi}, T.~D. and {Boyle}, M. and {Bozzi}, A.},
        title = "{GWTC-5.0: Population Properties of Merging Compact Binaries}",
      journal = {arXiv e-prints},
         year = 2026,
        month = may,
          eid = {arXiv:2605.27226},
        pages = {arXiv:2605.27226},
          doi = {10.48550/arXiv.2605.27226},
archivePrefix = {arXiv},
       eprint = {2605.27226},
 primaryClass = {astro-ph.HE},
       adsurl = {https://ui.adsabs.harvard.edu/abs/2026arXiv260527226T}
}

@ARTICLE{gwtc5_catalog,
       author = {{The LIGO Scientific Collaboration} and {the Virgo Collaboration} and {the KAGRA Collaboration} and {Abac}, A.~G. and {Abe}, A. and {Abouelfettouh}, I. and {Acernese}, F. and {Ackley}, K. and {Adam}, A. and {Adhicary}, S. and {Adhikari}, D. and {Adhikari}, R.~X. and {Adkins}, V.~K. and {Afroz}, S. and {Agapito}, A. and {Agarwal}, D. and {Agathos}, M. and {Aggarwal}, N. and {Aggarwal}, S. and {Aguiar}, O.~D. and {Ahrend}, I.-L. and {Aiello}, L. and {Ain}, A. and {Ajith}, P. and {Akutsu}, T. and {Albers}, L. and {Ali}, W. and {Al-Kershi}, S. and {Allene}, C. and {Allocca}, A. and {Al-Shammari}, S. and {Alvarez}, J.~A. and {Alvarez-Lopez}, S. and {Amar}, W. and {Amarasinghe}, O. and {Amato}, A. and {Amicucci}, F. and {Amra}, C. and {Anand}, A.~B. and {Anand}, C. and {Ananyeva}, A. and {Anderson}, S.~B. and {Anderson}, W.~G. and {Andia}, M. and {Ando}, M. and {Andrade-Oliveira}, F. and {Andr{\'e}s-Carcasona}, M. and {Andrey}, J.~L. and {Andri{\'c}}, T. and {Anglin}, J. and {Anna}, J. and {Antelis}, J.~M. and {Antier}, S. and {Aoki}, T. and {Aoumi}, M. and {Appavuravther}, E.~Z. and {Appelt}, E.~A. and {Appert}, S. and {Apple}, S.~K. and {Arai}, K. and {Araya}, A. and {Araya}, M.~C. and {Arca Sedda}, M. and {Arciprete}, F. and {Areeda}, J.~S. and {Aritomi}, N. and {Armato}, F. and {Armstrong}, S. and {Arnaud}, N. and {Arogeti}, M. and {Aronson}, S.~M. and {Ashton}, G. and {Aso}, Y. and {Asprea}, L. and {Assiduo}, M. and {Assis de Souza Melo}, S. and {Aston}, S.~M. and {Astone}, P. and {Aswathi}, P.~S. and {Attadio}, F. and {Aubin}, F. and {AultONeal}, K. and {Avallone}, G. and {Avdeev}, N. and {Avila}, E.~A. and {Babak}, S. and {Badger}, C. and {Bae}, S. and {Bagnasco}, S. and {Baimukhametova}, S. and {Baiotti}, L. and {Baka}, T. and {Baker}, K.~A. and {Baker}, T. and {Balbi}, G. and {Baldi}, G. and {Baldicchi}, N. and {Ball}, M. and {Ballardin}, G. and {Ballelli}, M. and {Ballmer}, S.~W. and {Banagiri}, S. and {Banerjee}, B. and {Bankar}, D. and {Baptiste}, T.~M. and {Baral}, P. and {Baratti}, M. and {Barayoga}, J.~C. and {Baric}, K. and {Barish}, B.~C. and {Barker}, D. and {Barman}, N. and {Barone}, F. and {Barr}, B. and {Barrios}, M. and {Barsotti}, L. and {Barsuglia}, M. and {Barta}, D. and {Barton}, M.~A. and {Bartos}, I. and {Basalaev}, A. and {Bassiri}, R. and {Basti}, A. and {Bawaj}, M. and {Bayley}, J.~C. and {Baylor}, A.~C. and {Baynard}, II, P.~A. and {Bazzan}, M. and {Bedakihale}, V.~M. and {Beirnaert}, F. and {Bejger}, M. and {Bell}, A.~S. and {Bellani}, C. and {Bellie}, D.~S. and {Beltran-Martinez}, D. and {Benedetti}, E. and {Benoit}, W. and {Bentara}, I. and {Ben Yaala}, M. and {Bera}, S. and {Bergamin}, F. and {Berger}, B.~K. and {Beroiz}, M. and {Berry}, C.~P.~L. and {Berry}, I. and {Bersanetti}, D. and {Bertheas}, T. and {Bertolini}, A. and {Betzwieser}, J. and {Beveridge}, D. and {Bevins}, N. and {Bezerra-Sobrinho}, J. and {Bhandare}, R. and {Bhatt}, R. and {Bhattacharjee}, A. and {Bhattacharjee}, D. and {Bhattacharyya}, S. and {Bhaumik}, S. and {Biancalana}, V. and {Bianchi}, F. and {Bilenko}, I.~A. and {Bilicki}, M. and {Billingsley}, G. and {Binetti}, A. and {Bini}, S. and {Biot}, S. and {Birnholtz}, O. and {Biscoveanu}, S. and {Bisht}, A. and {Bitossi}, M. and {Bizouard}, M.-A. and {Blaber}, S. and {Blackburn}, J.~K. and {Blagg}, L.~A. and {Blair}, C.~D. and {Blair}, D.~G. and {Bloch}, M. and {Bode}, N. and {Boettner}, N. and {Bogdan}, P. and {Boileau}, G. and {Boldrini}, M. and {Bolingbroke}, G.~N. and {Bonavena}, L.~D. and {Bonhomme}, V.~A. and {Bonilla}, E. and {Bonilla}, M.~S. and {Bonino}, A. and {Bonnand}, R. and {Borchers}, A. and {Borghi}, N. and {Boschi}, V. and {Bose}, S. and {Bossilkov}, V. and {Bothra}, Y. and {Boudon}, A. and {Boybeyi}, T.~D. and {Boyle}, M. and {Bozzi}, A. and {Bradaschia}, C.},
        title = "{GWTC-5.0: Observations from the Second Part of the Fourth LIGO-Virgo-KAGRA Observing Run and Updates to the Gravitational-Wave Transient Catalog}",
      journal = {arXiv e-prints},
         year = 2026,
        month = may,
          eid = {arXiv:2605.27225},
        pages = {arXiv:2605.27225},
          doi = {10.48550/arXiv.2605.27225},
archivePrefix = {arXiv},
       eprint = {2605.27225},
 primaryClass = {gr-qc},
       adsurl = {https://ui.adsabs.harvard.edu/abs/2026arXiv260527225T}
}

@ARTICLE{GW190814,
       author = {{Lyu}, F. and {Yuan}, L. and {Wu}, D.~H. and {Guo}, W.~H. and {Wang}, Y.~Z. and {Yi}, S.~X. and {Tang}, Q.~W. and {Hu}, R.-C. and {Zhu}, J.-P. and {Shu}, X.~W. and {Qin}, Y. and {Liang}, E.~W.},
        title = "{Revisiting the properties of GW190814 and its formation history}",
      journal = {\mnras},
         year = 2023,
        month = nov,
       volume = {525},
       number = {3},
        pages = {4321-4328},
          doi = {10.1093/mnras/stad2538},
archivePrefix = {arXiv},
       eprint = {2308.09893},
 primaryClass = {astro-ph.HE},
       adsurl = {https://ui.adsabs.harvard.edu/abs/2023MNRAS.525.4321L}
}

@ARTICLE{Mandel2021,
       author = {{Mandel}, Ilya and {Smith}, Rory J.~E.},
        title = "{GW200115: A Nonspinning Black Hole-Neutron Star Merger}",
      journal = {\apjl},
         year = 2021,
        month = nov,
       volume = {922},
       number = {1},
          eid = {L14},
        pages = {L14},
          doi = {10.3847/2041-8213/ac35dd},
archivePrefix = {arXiv},
       eprint = {2109.14759},
 primaryClass = {astro-ph.HE},
       adsurl = {https://ui.adsabs.harvard.edu/abs/2021ApJ...922L..14M}
}

@ARTICLE{Xue2025,
       author = {{Xue}, Ya-Wen and {Qin}, Ying and {Yuan}, Liang and {Guo}, Wei-Hua and {Li}, Jun-Qian and {Zhang}, Yu-Qing and {Wang}, Zi-Yuan and {Wu}, Dong-Hong},
        title = "{Astrophysical Priors on the Properties of Observed Black Hole─Neutron Star Mergers}",
      journal = {Research in Astronomy and Astrophysics},
         year = 2025,
        month = apr,
       volume = {25},
       number = {4},
          eid = {045009},
        pages = {045009},
          doi = {10.1088/1674-4527/adc64f},
       adsurl = {https://ui.adsabs.harvard.edu/abs/2025RAA....25d5009X}
}

@ARTICLE{Mandel2020,
       author = {{Mandel}, Ilya and {Fragos}, Tassos},
        title = "{An Alternative Interpretation of GW190412 as a Binary Black Hole Merger with a Rapidly Spinning Secondary}",
      journal = {\apjl},
         year = 2020,
        month = jun,
       volume = {895},
       number = {2},
          eid = {L28},
        pages = {L28},
          doi = {10.3847/2041-8213/ab8e41},
archivePrefix = {arXiv},
       eprint = {2004.09288},
 primaryClass = {astro-ph.HE},
       adsurl = {https://ui.adsabs.harvard.edu/abs/2020ApJ...895L..28M}
}

@ARTICLE{Abbott2021,
       author = {{Abbott}, R. and {Abbott}, T.~D. and {Abraham}, S. and {Acernese}, F. and {Ackley}, K. and {Adams}, A. and {Adams}, C. and {Adhikari}, R.~X. and {Adya}, V.~B. and {Affeldt}, C. and {Agathos}, M. and {Agatsuma}, K. and {Aggarwal}, N. and {Aguiar}, O.~D. and {Aiello}, L. and {Ain}, A. and {Ajith}, P. and {Akcay}, S. and {Allen}, G. and {Allocca}, A. and {Altin}, P.~A. and {Amato}, A. and {Anand}, S. and {Ananyeva}, A. and {Anderson}, S.~B. and {Anderson}, W.~G. and {Angelova}, S.~V. and {Ansoldi}, S. and {Antelis}, J.~M. and {Antier}, S. and {Appert}, S. and {Arai}, K. and {Araya}, M.~C. and {Areeda}, J.~S. and {Ar{\`e}ne}, M. and {Arnaud}, N. and {Aronson}, S.~M. and {Arun}, K.~G. and {Asali}, Y. and {Ascenzi}, S. and {Ashton}, G. and {Aston}, S.~M. and {Astone}, P. and {Aubin}, F. and {Aufmuth}, P. and {AultONeal}, K. and {Austin}, C. and {Avendano}, V. and {Babak}, S. and {Badaracco}, F. and {Bader}, M.~K.~M. and {Bae}, S. and {Baer}, A.~M. and {Bagnasco}, S. and {Baird}, J. and {Ball}, M. and {Ballardin}, G. and {Ballmer}, S.~W. and {Bals}, A. and {Balsamo}, A. and {Baltus}, G. and {Banagiri}, S. and {Bankar}, D. and {Bankar}, R.~S. and {Barayoga}, J.~C. and {Barbieri}, C. and {Barish}, B.~C. and {Barker}, D. and {Barneo}, P. and {Barnum}, S. and {Barone}, F. and {Barr}, B. and {Barsotti}, L. and {Barsuglia}, M. and {Barta}, D. and {Bartlett}, J. and {Bartos}, I. and {Bassiri}, R. and {Basti}, A. and {Bawaj}, M. and {Bayley}, J.~C. and {Bazzan}, M. and {Becher}, B.~R. and {B{\'e}csy}, B. and {Bedakihale}, V.~M. and {Bejger}, M. and {Belahcene}, I. and {Beniwal}, D. and {Benjamin}, M.~G. and {Bennett}, T.~F. and {Bentley}, J.~D. and {Bergamin}, F. and {Berger}, B.~K. and {Bergmann}, G. and {Bernuzzi}, S. and {Berry}, C.~P.~L. and {Bersanetti}, D. and {Bertolini}, A. and {Betzwieser}, J. and {Bhandare}, R. and {Bhandari}, A.~V. and {Bhattacharjee}, D. and {Bidler}, J. and {Bilenko}, I.~A. and {Billingsley}, G. and {Birney}, R. and {Birnholtz}, O. and {Biscans}, S. and {Bischi}, M. and {Biscoveanu}, S. and {Bisht}, A. and {Bitossi}, M. and {Bizouard}, M.-A. and {Blackburn}, J.~K. and {Blackman}, J. and {Blair}, C.~D. and {Blair}, D.~G. and {Blair}, R.~M. and {Blanch}, O. and {Bobba}, F. and {Bode}, N. and {Boer}, M. and {Boetzel}, Y. and {Bogaert}, G. and {Boldrini}, M. and {Bondu}, F. and {Bonilla}, E. and {Bonnand}, R. and {Booker}, P. and {Boom}, B.~A. and {Bork}, R. and {Boschi}, V. and {Bose}, S. and {Bossilkov}, V. and {Boudart}, V. and {Bouffanais}, Y. and {Bozzi}, A. and {Bradaschia}, C. and {Brady}, P.~R. and {Bramley}, A. and {Branchesi}, M. and {Brau}, J.~E. and {Breschi}, M. and {Briant}, T. and {Briggs}, J.~H. and {Brighenti}, F. and {Brillet}, A. and {Brinkmann}, M. and {Brockill}, P. and {Brooks}, A.~F. and {Brooks}, J. and {Brown}, D.~D. and {Brunett}, S. and {Bruno}, G. and {Bruntz}, R. and {Buikema}, A. and {Bulik}, T. and {Bulten}, H.~J. and {Buonanno}, A. and {Buscicchio}, R. and {Buskulic}, D. and {Byer}, R.~L. and {Cabero}, M. and {Cadonati}, L. and {Caesar}, M. and {Cagnoli}, G. and {Cahillane}, C. and {Calder{\'o}n Bustillo}, J. and {Callaghan}, J.~D. and {Callister}, T.~A. and {Calloni}, E. and {Camp}, J.~B. and {Canepa}, M. and {Cannon}, K.~C. and {Cao}, H. and {Cao}, J. and {Carapella}, G. and {Carbognani}, F. and {Carney}, M.~F. and {Carpinelli}, M. and {Carullo}, G. and {Carver}, T.~L. and {Casanueva Diaz}, J. and {Casentini}, C. and {Caudill}, S. and {Cavagli{\`a}}, M. and {Cavalier}, F. and {Cavalieri}, R. and {Cella}, G. and {Cerd{\'a}-Dur{\'a}n}, P. and {Cesarini}, E. and {Chaibi}, W. and {Chakravarti}, K. and {Chan}, C.-L. and {Chan}, C. and {Chandra}, K. and {Chanial}, P. and {Chao}, S. and {Charlton}, P. and {Chase}, E.~A.},
        title = "{GWTC-2: Compact Binary Coalescences Observed by LIGO and Virgo during the First Half of the Third Observing Run}",
      journal = {Physical Review X},
         year = 2021,
        month = apr,
       volume = {11},
       number = {2},
          eid = {021053},
        pages = {021053},
          doi = {10.1103/PhysRevX.11.021053},
archivePrefix = {arXiv},
       eprint = {2010.14527},
 primaryClass = {gr-qc},
       adsurl = {https://ui.adsabs.harvard.edu/abs/2021PhRvX..11b1053A}
}

@ARTICLE{Abbott2024,
       author = {{Abbott}, R. and {Abbott}, T.~D. and {Acernese}, F. and {Ackley}, K. and {Adams}, C. and {Adhikari}, N. and {Adhikari}, R.~X. and {Adya}, V.~B. and {Affeldt}, C. and {Agarwal}, D. and {Agathos}, M. and {Agatsuma}, K. and {Aggarwal}, N. and {Aguiar}, O.~D. and {Aiello}, L. and {Ain}, A. and {Ajith}, P. and {Albanesi}, S. and {Allocca}, A. and {Altin}, P.~A. and {Amato}, A. and {Anand}, C. and {Anand}, S. and {Ananyeva}, A. and {Anderson}, S.~B. and {Anderson}, W.~G. and {Andrade}, T. and {Andres}, N. and {Andri{\'c}}, T. and {Angelova}, S.~V. and {Ansoldi}, S. and {Antelis}, J.~M. and {Antier}, S. and {Appert}, S. and {Arai}, K. and {Araya}, M.~C. and {Areeda}, J.~S. and {Ar{\`e}ne}, M. and {Arnaud}, N. and {Aronson}, S.~M. and {Arun}, K.~G. and {Asali}, Y. and {Ashton}, G. and {Assiduo}, M. and {Aston}, S.~M. and {Astone}, P. and {Aubin}, F. and {Austin}, C. and {Babak}, S. and {Badaracco}, F. and {Bader}, M.~K.~M. and {Badger}, C. and {Bae}, S. and {Baer}, A.~M. and {Bagnasco}, S. and {Bai}, Y. and {Baird}, J. and {Ball}, M. and {Ballardin}, G. and {Ballmer}, S.~W. and {Balsamo}, A. and {Baltus}, G. and {Banagiri}, S. and {Bankar}, D. and {Barayoga}, J.~C. and {Barbieri}, C. and {Barish}, B.~C. and {Barker}, D. and {Barneo}, P. and {Barone}, F. and {Barr}, B. and {Barsotti}, L. and {Barsuglia}, M. and {Barta}, D. and {Bartlett}, J. and {Barton}, M.~A. and {Bartos}, I. and {Bassiri}, R. and {Basti}, A. and {Bawaj}, M. and {Bayley}, J.~C. and {Baylor}, A.~C. and {Bazzan}, M. and {B{\'e}csy}, B. and {Bedakihale}, V.~M. and {Bejger}, M. and {Belahcene}, I. and {Benedetto}, V. and {Beniwal}, D. and {Bennett}, T.~F. and {Bentley}, J.~D. and {BenYaala}, M. and {Bergamin}, F. and {Berger}, B.~K. and {Bernuzzi}, S. and {Berry}, C.~P.~L. and {Bersanetti}, D. and {Bertolini}, A. and {Betzwieser}, J. and {Beveridge}, D. and {Bhandare}, R. and {Bhardwaj}, U. and {Bhattacharjee}, D. and {Bhaumik}, S. and {Bilenko}, I.~A. and {Billingsley}, G. and {Bini}, S. and {Birney}, R. and {Birnholtz}, O. and {Biscans}, S. and {Bischi}, M. and {Biscoveanu}, S. and {Bisht}, A. and {Biswas}, B. and {Bitossi}, M. and {Bizouard}, M.-A. and {Blackburn}, J.~K. and {Blair}, C.~D. and {Blair}, D.~G. and {Blair}, R.~M. and {Bobba}, F. and {Bode}, N. and {Boer}, M. and {Bogaert}, G. and {Boldrini}, M. and {Bonavena}, L.~D. and {Bondu}, F. and {Bonilla}, E. and {Bonnand}, R. and {Booker}, P. and {Boom}, B.~A. and {Bork}, R. and {Boschi}, V. and {Bose}, N. and {Bose}, S. and {Bossilkov}, V. and {Boudart}, V. and {Bouffanais}, Y. and {Bozzi}, A. and {Bradaschia}, C. and {Brady}, P.~R. and {Bramley}, A. and {Branch}, A. and {Branchesi}, M. and {Brau}, J.~E. and {Breschi}, M. and {Briant}, T. and {Briggs}, J.~H. and {Brillet}, A. and {Brinkmann}, M. and {Brockill}, P. and {Brooks}, A.~F. and {Brooks}, J. and {Brown}, D.~D. and {Brunett}, S. and {Bruno}, G. and {Bruntz}, R. and {Bryant}, J. and {Bulik}, T. and {Bulten}, H.~J. and {Buonanno}, A. and {Buscicchio}, R. and {Buskulic}, D. and {Buy}, C. and {Byer}, R.~L. and {Cadonati}, L. and {Cagnoli}, G. and {Cahillane}, C. and {Bustillo}, J. Calder{\'o}n and {Callaghan}, J.~D. and {Callister}, T.~A. and {Calloni}, E. and {Cameron}, J. and {Camp}, J.~B. and {Canepa}, M. and {Canevarolo}, S. and {Cannavacciuolo}, M. and {Cannon}, K.~C. and {Cao}, H. and {Capote}, E. and {Carapella}, G. and {Carbognani}, F. and {Carlin}, J.~B. and {Carney}, M.~F. and {Carpinelli}, M. and {Carrillo}, G. and {Carullo}, G. and {Carver}, T.~L. and {Diaz}, J. Casanueva and {Casentini}, C. and {Castaldi}, G. and {Caudill}, S. and {Cavagli{\`a}}, M. and {Cavalier}, F. and {Cavalieri}, R. and {Ceasar}, M. and {Cella}, G. and {Cerd{\'a}-Dur{\'a}n}, P. and {Cesarini}, E. and {Chaibi}, W.},
        title = "{GWTC-2.1: Deep extended catalog of compact binary coalescences observed by LIGO and Virgo during the first half of the third observing run}",
      journal = {\prd},
         year = 2024,
        month = jan,
       volume = {109},
       number = {2},
          eid = {022001},
        pages = {022001},
          doi = {10.1103/PhysRevD.109.022001},
archivePrefix = {arXiv},
       eprint = {2108.01045},
 primaryClass = {gr-qc},
       adsurl = {https://ui.adsabs.harvard.edu/abs/2024PhRvD.109b2001A}
}

@ARTICLE{Olejak2021,
       author = {{Olejak}, A. and {Belczynski}, K.},
        title = "{The Implications of High Black Hole Spins for the Origin of Binary Black Hole Mergers}",
      journal = {\apjl},
         year = 2021,
        month = nov,
       volume = {921},
       number = {1},
          eid = {L2},
        pages = {L2},
          doi = {10.3847/2041-8213/ac2f48},
archivePrefix = {arXiv},
       eprint = {2109.06872},
 primaryClass = {astro-ph.HE},
       adsurl = {https://ui.adsabs.harvard.edu/abs/2021ApJ...921L...2O}
}

@ARTICLE{Hu2022,
       author = {{Hu}, Rui-Chong and {Zhu}, Jin-Ping and {Qin}, Ying and {Zhang}, Bing and {Liang}, En-Wei and {Shao}, Yong},
        title = "{A Channel to Form Fast-spinning Black Hole-Neutron Star Binary Mergers as Multimessenger Sources}",
      journal = {\apj},
         year = 2022,
        month = apr,
       volume = {928},
       number = {2},
          eid = {163},
        pages = {163},
          doi = {10.3847/1538-4357/ac573f},
archivePrefix = {arXiv},
       eprint = {2201.09549},
 primaryClass = {astro-ph.HE},
       adsurl = {https://ui.adsabs.harvard.edu/abs/2022ApJ...928..163H}
}

@ARTICLE{Qin2018,
       author = {{Qin}, Y. and {Fragos}, T. and {Meynet}, G. and {Andrews}, J. and {S{\o}rensen}, M. and {Song}, H.~F.},
        title = "{The spin of the second-born black hole in coalescing binary black holes}",
      journal = {\aap},
         year = 2018,
        month = aug,
       volume = {616},
          eid = {A28},
        pages = {A28},
          doi = {10.1051/0004-6361/201832839},
archivePrefix = {arXiv},
       eprint = {1802.05738},
 primaryClass = {astro-ph.SR},
       adsurl = {https://ui.adsabs.harvard.edu/abs/2018A&A...616A..28Q}
}

@ARTICLE{fuller2019,
       author = {{Fuller}, Jim and {Ma}, Linhao},
        title = "{Most Black Holes Are Born Very Slowly Rotating}",
      journal = {\apjl},
         year = 2019,
        month = aug,
       volume = {881},
       number = {1},
          eid = {L1},
        pages = {L1},
          doi = {10.3847/2041-8213/ab339b},
archivePrefix = {arXiv},
       eprint = {1907.03714},
 primaryClass = {astro-ph.SR},
       adsurl = {https://ui.adsabs.harvard.edu/abs/2019ApJ...881L...1F}
}

@ARTICLE{Belczynski2016,
       author = {{Belczynski}, Krzysztof and {Repetto}, Serena and {Holz}, Daniel E. and {O'Shaughnessy}, Richard and {Bulik}, Tomasz and {Berti}, Emanuele and {Fryer}, Christopher and {Dominik}, Michal},
        title = "{Compact Binary Merger Rates: Comparison with LIGO/Virgo Upper Limits}",
      journal = {\apj},
         year = 2016,
        month = mar,
       volume = {819},
       number = {2},
          eid = {108},
        pages = {108},
          doi = {10.3847/0004-637X/819/2/108},
archivePrefix = {arXiv},
       eprint = {1510.04615},
 primaryClass = {astro-ph.HE},
       adsurl = {https://ui.adsabs.harvard.edu/abs/2016ApJ...819..108B}
}

@ARTICLE{Bavera2020,
       author = {{Bavera}, Simone S. and {Fragos}, Tassos and {Qin}, Ying and {Zapartas}, Emmanouil and {Neijssel}, Coenraad J. and {Mandel}, Ilya and {Batta}, Aldo and {Gaebel}, Sebastian M. and {Kimball}, Chase and {Stevenson}, Simon},
        title = "{The origin of spin in binary black holes. Predicting the distributions of the main observables of Advanced LIGO}",
      journal = {\aap},
         year = 2020,
        month = mar,
       volume = {635},
          eid = {A97},
        pages = {A97},
          doi = {10.1051/0004-6361/201936204},
archivePrefix = {arXiv},
       eprint = {1906.12257},
 primaryClass = {astro-ph.HE},
       adsurl = {https://ui.adsabs.harvard.edu/abs/2020A&A...635A..97B}
}

@ARTICLE{Abbott2020,
       author = {{Abbott}, R. and {Abbott}, T.~D. and {Abraham}, S. and {Acernese}, F. and {Ackley}, K. and {Adams}, C. and {Adhikari}, R.~X. and {Adya}, V.~B. and {Affeldt}, C. and {Agathos}, M. and {Agatsuma}, K. and {Aggarwal}, N. and {Aguiar}, O.~D. and {Aich}, A. and {Aiello}, L. and {Ain}, A. and {Ajith}, P. and {Akcay}, S. and {Allen}, G. and {Allocca}, A. and {Altin}, P.~A. and {Amato}, A. and {Anand}, S. and {Ananyeva}, A. and {Anderson}, S.~B. and {Anderson}, W.~G. and {Angelova}, S.~V. and {Ansoldi}, S. and {Antier}, S. and {Appert}, S. and {Arai}, K. and {Araya}, M.~C. and {Areeda}, J.~S. and {Ar{\`e}ne}, M. and {Arnaud}, N. and {Aronson}, S.~M. and {Arun}, K.~G. and {Asali}, Y. and {Ascenzi}, S. and {Ashton}, G. and {Aston}, S.~M. and {Astone}, P. and {Aubin}, F. and {Aufmuth}, P. and {AultONeal}, K. and {Austin}, C. and {Avendano}, V. and {Babak}, S. and {Bacon}, P. and {Badaracco}, F. and {Bader}, M.~K.~M. and {Bae}, S. and {Baer}, A.~M. and {Baird}, J. and {Baldaccini}, F. and {Ballardin}, G. and {Ballmer}, S.~W. and {Bals}, A. and {Balsamo}, A. and {Baltus}, G. and {Banagiri}, S. and {Bankar}, D. and {Bankar}, R.~S. and {Barayoga}, J.~C. and {Barbieri}, C. and {Barish}, B.~C. and {Barker}, D. and {Barkett}, K. and {Barneo}, P. and {Barone}, F. and {Barr}, B. and {Barsotti}, L. and {Barsuglia}, M. and {Barta}, D. and {Bartlett}, J. and {Bartos}, I. and {Bassiri}, R. and {Basti}, A. and {Bawaj}, M. and {Bayley}, J.~C. and {Bazzan}, M. and {B{\'e}csy}, B. and {Bejger}, M. and {Belahcene}, I. and {Bell}, A.~S. and {Beniwal}, D. and {Benjamin}, M.~G. and {Benkel}, R. and {Bentley}, J.~D. and {Bergamin}, F. and {Berger}, B.~K. and {Bergmann}, G. and {Bernuzzi}, S. and {Berry}, C.~P.~L. and {Bersanetti}, D. and {Bertolini}, A. and {Betzwieser}, J. and {Bhandare}, R. and {Bhandari}, A.~V. and {Bidler}, J. and {Biggs}, E. and {Bilenko}, I.~A. and {Billingsley}, G. and {Birney}, R. and {Birnholtz}, O. and {Biscans}, S. and {Bischi}, M. and {Biscoveanu}, S. and {Bisht}, A. and {Bissenbayeva}, G. and {Bitossi}, M. and {Bizouard}, M.~A. and {Blackburn}, J.~K. and {Blackman}, J. and {Blair}, C.~D. and {Blair}, D.~G. and {Blair}, R.~M. and {Bobba}, F. and {Bode}, N. and {Boer}, M. and {Boetzel}, Y. and {Bogaert}, G. and {Bondu}, F. and {Bonilla}, E. and {Bonnand}, R. and {Booker}, P. and {Boom}, B.~A. and {Bork}, R. and {Boschi}, V. and {Bose}, S. and {Bossilkov}, V. and {Bosveld}, J. and {Bouffanais}, Y. and {Bozzi}, A. and {Bradaschia}, C. and {Brady}, P.~R. and {Bramley}, A. and {Branchesi}, M. and {Brau}, J.~E. and {Breschi}, M. and {Briant}, T. and {Briggs}, J.~H. and {Brighenti}, F. and {Brillet}, A. and {Brinkmann}, M. and {Brito}, R. and {Brockill}, P. and {Brooks}, A.~F. and {Brooks}, J. and {Brown}, D.~D. and {Brunett}, S. and {Bruno}, G. and {Bruntz}, R. and {Buikema}, A. and {Bulik}, T. and {Bulten}, H.~J. and {Buonanno}, A. and {Buskulic}, D. and {Byer}, R.~L. and {Cabero}, M. and {Cadonati}, L. and {Cagnoli}, G. and {Cahillane}, C. and {Calder{\'o}n Bustillo}, J. and {Callaghan}, J.~D. and {Callister}, T.~A. and {Calloni}, E. and {Camp}, J.~B. and {Canepa}, M. and {Cannon}, K.~C. and {Cao}, H. and {Cao}, J. and {Carapella}, G. and {Carbognani}, F. and {Caride}, S. and {Carney}, M.~F. and {Carullo}, G. and {Casanueva Diaz}, J. and {Casentini}, C. and {Casta{\~n}eda}, J. and {Caudill}, S. and {Cavagli{\`a}}, M. and {Cavalier}, F. and {Cavalieri}, R. and {Cella}, G. and {Cerd{\'a}-Dur{\'a}n}, P. and {Cesarini}, E. and {Chaibi}, O. and {Chakravarti}, K. and {Chan}, C. and {Chan}, M. and {Chao}, S. and {Charlton}, P. and {Chase}, E.~A. and {Chassande-Mottin}, E. and {Chatterjee}, D. and {Chaturvedi}, M. and {Chatziioannou}, K. and {Chen}, H.~Y. and {Chen}, X.},
        title = "{GW190412: Observation of a binary-black-hole coalescence with asymmetric masses}",
      journal = {\prd},
         year = 2020,
        month = aug,
       volume = {102},
       number = {4},
          eid = {043015},
        pages = {043015},
          doi = {10.1103/PhysRevD.102.043015},
archivePrefix = {arXiv},
       eprint = {2004.08342},
 primaryClass = {astro-ph.HE},
       adsurl = {https://ui.adsabs.harvard.edu/abs/2020PhRvD.102d3015A}
}

@ARTICLE{Colleoni2025,
       author = {{Colleoni}, Marta and {Ramis Vidal}, Felip A. and {Garc{\'\i}a-Quir{\'o}s}, Cecilio and {Ak{\c{c}}ay}, Sarp and {Bera}, Sayantani},
        title = "{Fast frequency-domain gravitational waveforms for precessing binaries with a new twist}",
      journal = {\prd},
         year = 2025,
        month = may,
       volume = {111},
       number = {10},
          eid = {104019},
        pages = {104019},
          doi = {10.1103/PhysRevD.111.104019},
archivePrefix = {arXiv},
       eprint = {2412.16721},
 primaryClass = {gr-qc},
       adsurl = {https://ui.adsabs.harvard.edu/abs/2025PhRvD.111j4019C}
}

@ARTICLE{Broekgaarden2022,
       author = {{Broekgaarden}, Floor S. and {Stevenson}, Simon and {Thrane}, Eric},
        title = "{Signatures of Mass Ratio Reversal in Gravitational Waves from Merging Binary Black Holes}",
      journal = {\apj},
         year = 2022,
        month = oct,
       volume = {938},
       number = {1},
          eid = {45},
        pages = {45},
          doi = {10.3847/1538-4357/ac8879},
archivePrefix = {arXiv},
       eprint = {2205.01693},
 primaryClass = {astro-ph.HE},
       adsurl = {https://ui.adsabs.harvard.edu/abs/2022ApJ...938...45B}
}

@ARTICLE{Hu2026,
       author = {{Hu}, Rui-Chong and {Qin}, Ying and {Zhang}, Bing},
        title = "{Mass-Ratio Reversal as an Alternative to Hierarchical Mergers for GW241011}",
      journal = {arXiv e-prints},
         year = 2026,
        month = jun,
          eid = {arXiv:2606.27852},
        pages = {arXiv:2606.27852},
          doi = {10.48550/arXiv.2606.27852},
archivePrefix = {arXiv},
       eprint = {2606.27852},
 primaryClass = {astro-ph.HE},
       adsurl = {https://ui.adsabs.harvard.edu/abs/2026arXiv260627852H}
}

@ARTICLE{GW241011,
       author = {{Abac}, A.~G. and {Abouelfettouh}, I. and {Acernese}, F. and {Ackley}, K. and {Adamcewicz}, C. and {Adhicary}, S. and {Adhikari}, D. and {Adhikari}, N. and {Adhikari}, R.~X. and {Adkins}, V.~K. and {Afroz}, S. and {Agapito}, A. and {Agarwal}, D. and {Agathos}, M. and {Aggarwal}, N. and {Aggarwal}, S. and {Aguiar}, O.~D. and {Ahrend}, I.-L. and {Aiello}, L. and {Ain}, A. and {Ajith}, P. and {Akutsu}, T. and {Albanesi}, S. and {Ali}, W. and {Al-Kershi}, S. and {All{\'e}n{\'e}}, C. and {Allocca}, A. and {Al-Shammari}, S. and {Altin}, P.~A. and {Alvarez-Lopez}, S. and {Amar}, W. and {Amarasinghe}, O. and {Amato}, A. and {Amicucci}, F. and {Amra}, C. and {Ananyeva}, A. and {Anderson}, S.~B. and {Anderson}, W.~G. and {Andia}, M. and {Ando}, M. and {Andr{\'e}s-Carcasona}, M. and {Andri{\'c}}, T. and {Anglin}, J. and {Ansoldi}, S. and {Antelis}, J.~M. and {Antier}, S. and {Antonini}, F. and {Aoumi}, M. and {Appavuravther}, E.~Z. and {Appert}, S. and {Apple}, S.~K. and {Arai}, K. and {Ara{\'u}jo-{\'A}lvarez}, C. and {Araya}, A. and {Araya}, M.~C. and {Arca Sedda}, M. and {Areeda}, J.~S. and {Aritomi}, N. and {Armato}, F. and {Armstrong}, S. and {Arnaud}, N. and {Arogeti}, M. and {Aronson}, S.~M. and {Arun}, K.~G. and {Ashton}, G. and {Aso}, Y. and {Asprea}, L. and {Assiduo}, M. and {Assis de Souza Melo}, S. and {Aston}, S.~M. and {Astone}, P. and {Aswathi}, P.~S. and {Attadio}, F. and {Aubin}, F. and {Aultoneal}, K. and {Avallone}, G. and {Avila}, E.~A. and {Babak}, S. and {Badger}, C. and {Bae}, S. and {Bagnasco}, S. and {Baiotti}, L. and {Bajpai}, R. and {Baka}, T. and {Baker}, A.~M. and {Baker}, K.~A. and {Baker}, T. and {Baldi}, G. and {Baldicchi}, N. and {Ball}, M. and {Ballardin}, G. and {Ballmer}, S.~W. and {Banagiri}, S. and {Banerjee}, B. and {Bankar}, D. and {Baptiste}, T.~M. and {Baral}, P. and {Baratti}, M. and {Barayoga}, J.~C. and {Barish}, B.~C. and {Barker}, D. and {Barman}, N. and {Barneo}, P. and {Barone}, F. and {Barr}, B. and {Barsotti}, L. and {Barsuglia}, M. and {Barta}, D. and {Bartoletti}, A.~M. and {Barton}, M.~A. and {Bartos}, I. and {Basalaev}, A. and {Bassiri}, R. and {Basti}, A. and {Bawaj}, M. and {Baxi}, P. and {Bayley}, J.~C. and {Baylor}, A.~C. and {Baynard}, II, P.~A. and {Bazzan}, M. and {Bedakihale}, V.~M. and {Beirnaert}, F. and {Bejger}, M. and {Belardinelli}, D. and {Bell}, A.~S. and {Bellie}, D.~S. and {Bellizzi}, L. and {Benoit}, W. and {Bentara}, I. and {Bentley}, J.~D. and {Ben Yaala}, M. and {Bera}, S. and {Bergamin}, F. and {Berger}, B.~K. and {Bernuzzi}, S. and {Beroiz}, M. and {Berry}, C.~P.~L. and {Bersanetti}, D. and {Bertheas}, T. and {Bertolini}, A. and {Betzwieser}, J. and {Beveridge}, D. and {Bevilacqua}, G. and {Bevins}, N. and {Bhandare}, R. and {Bhatt}, R. and {Bhattacharjee}, D. and {Bhattacharyya}, S. and {Bhaumik}, S. and {Biancalana}, V. and {Bianchi}, A. and {Bilenko}, I.~A. and {Billingsley}, G. and {Binetti}, A. and {Bini}, S. and {Binu}, C. and {Biot}, S. and {Birnholtz}, O. and {Biscoveanu}, S. and {Bisht}, A. and {Bitossi}, M. and {Bizouard}, M.-A. and {Blaber}, S. and {Blackburn}, J.~K. and {Blagg}, L.~A. and {Blair}, C.~D. and {Blair}, D.~G. and {Bode}, N. and {Boettner}, N. and {Boileau}, G. and {Boldrini}, M. and {Bolingbroke}, G.~N. and {Bolliand}, A. and {Bonavena}, L.~D. and {Bondarescu}, R. and {Bondu}, F. and {Bonilla}, E. and {Bonilla}, M.~S. and {Bonino}, A. and {Bonnand}, R. and {Borchers}, A. and {Borhanian}, S. and {Boschi}, V. and {Bose}, S. and {Bossilkov}, V. and {Bothra}, Y. and {Boudon}, A. and {Bourg}, L. and {Boyle}, M. and {Bozzi}, A. and {Bradaschia}, C. and {Brady}, P.~R. and {Branch}, A. and {Branchesi}, M. and {Braun}, I. and {Briant}, T. and {Brillet}, A. and {Brinkmann}, M. and {Brockill}, P. and {Brockmueller}, E.},
        title = "{GW241011 and GW241110: Exploring Binary Formation and Fundamental Physics with Asymmetric, High-spin Black Hole Coalescences}",
      journal = {\apjl},
         year = 2025,
        month = nov,
       volume = {993},
       number = {1},
          eid = {L21},
        pages = {L21},
          doi = {10.3847/2041-8213/ae0d54},
archivePrefix = {arXiv},
       eprint = {2510.26931},
 primaryClass = {astro-ph.HE},
       adsurl = {https://ui.adsabs.harvard.edu/abs/2025ApJ...993L..21A}
}

@article{Thrane2019,
  author  = {Thrane, Eric and Talbot, Colm},
  title   = {An Introduction to Bayesian Inference in Gravitational-Wave
             Astronomy: Parameter Estimation, Model Selection, and
             Hierarchical Models},
  journal = {Publications of the Astronomical Society of Australia},
  year    = {2019},
  volume  = {36},
  pages   = {e010},
  doi     = {10.1017/pasa.2019.2}
}

@ARTICLE{Qin2022,
       author = {{Qin}, Ying and {Wang}, Yuan-Zhu and {Bavera}, Simone S. and {Wu}, Shichao and {Meynet}, Georges and {Wang}, Yi-Ying and {Hu}, Rui-Chong and {Zhu}, Jin-Ping and {Wu}, Dong-Hong and {Shu}, Xin-Wen and {Peng}, Fang-Kun and {Song}, Han-Feng and {Wei}, Da-Ming},
        title = "{Searching for Candidates of Coalescing Binary Black Holes Formed through Chemically Homogeneous Evolution in GWTC-3}",
      journal = {\apj},
         year = 2022,
        month = dec,
       volume = {941},
       number = {2},
          eid = {179},
        pages = {179},
          doi = {10.3847/1538-4357/aca40c},
archivePrefix = {arXiv},
       eprint = {2211.05945},
 primaryClass = {astro-ph.HE},
       adsurl = {https://ui.adsabs.harvard.edu/abs/2022ApJ...941..179Q}
}

@ARTICLE{Detmers2008,
       author = {{Detmers}, R.~G. and {Langer}, N. and {Podsiadlowski}, Ph. and {Izzard}, R.~G.},
        title = "{Gamma-ray bursts from tidally spun-up Wolf-Rayet stars?}",
      journal = {\aap},
         year = 2008,
        month = jun,
       volume = {484},
       number = {3},
        pages = {831-839},
          doi = {10.1051/0004-6361:200809371},
archivePrefix = {arXiv},
       eprint = {0804.0014},
 primaryClass = {astro-ph},
       adsurl = {https://ui.adsabs.harvard.edu/abs/2008A&A...484..831D}
}
\appendix


\label{lastpage}
\end{document}